\documentclass[twocolumn,showpacs,nofootinbib,preprintnumbers,amsmath,amssymb,showkeys,superscriptaddress]{revtex4-1}
\usepackage{graphicx}
\usepackage{amsmath}
\usepackage{amssymb}
\usepackage{dcolumn}
\usepackage{bm}
\usepackage{color}
\usepackage{dsfont}
\usepackage{epsfig}
\usepackage{setspace}

\newcommand \beq{\begin{eqnarray}}
\newcommand \eeq{\end{eqnarray}}

\newcommand \be{\begin{equation}}
\newcommand \ee{\end{equation}}

\newcommand{\rt}[1]{{}}

\newcommand\eqn[1]{(\ref{#1})}      
\newcommand\Eqn[1]{Eq.~(\ref{#1})}  
\newcommand\Fig[1]{Fig.~\ref{#1}}  

\begin{document}
\allowdisplaybreaks

\title{Localized rainbows in the QCD phase diagram}

	\author{J. Maelger}
	\affiliation{Centre de Physique Th\'eorique (CPHT), CNRS, Ecole Polytechnique, Institut Polytechnique de Paris,\\ Route de Saclay, F-91128 Palaiseau, France.\vspace{.1cm}}
	\affiliation{Astro-Particule et Cosmologie (APC), CNRS UMR 7164, Universit\'e Paris Diderot,\\ 10, rue Alice Domon et L\'eonie Duquet, 75205 Paris Cedex 13, France.\vspace{.1cm}}
	
	\author{U. Reinosa}%
	\affiliation{Centre de Physique Th\'eorique (CPHT), CNRS, Ecole Polytechnique, Institut Polytechnique de Paris,\\ Route de Saclay, F-91128 Palaiseau, France.\vspace{.1cm}}%
	
	\author{J. Serreau}
	\affiliation{Astro-Particule et Cosmologie (APC), CNRS UMR 7164, Universit\'e Paris Diderot,\\ 10, rue Alice Domon et L\'eonie Duquet, 75205 Paris Cedex 13, France.\vspace{.1cm}}

	\date{\today}

\begin{abstract}
We study the phase diagram of strongly interacting matter with light quarks using a recently proposed, small parameter approach to infrared QCD in the Landau gauge. This is based on an expansion with respect to both the inverse number of colors and the pure Yang-Mills coupling in the presence of a Curci-Ferrari mass term. At leading order, this leads to the well-known rainbow equation for the quark propagator with a massive gluon propagator. We solve the latter at nonzero temperature and chemical potential using a simple semi-analytic approximation known to capture the essence of chiral symmetry breaking in the vacuum. In the chiral limit, we find a tricritical point which becomes a critical endpoint in the presence of a nonzero bare quark mass, in agreement with the results of nonperturbative functional methods and model calculations. This supports the view that the present approach allows for a systematic study of the QCD phase diagram in a controlled expansion scheme. 
\end{abstract}

\maketitle

\section{Introduction}
Hadronic matter is expected to present a rich phase structure when submitted to sufficiently high energy and baryonic densities, large magnetic fields, etc., as encountered in various environments such as the early universe, ultradense astrophysical objects, or relativistic heavy ion collisions in the laboratory \cite{Fukushima:2010bq,Weissenborn:2011qu,Borsanyi:2016ksw}. Unravelling the phase diagram of quantum chromodynamics (QCD) at nonzero temperature $T$ and baryonic chemical potential  $\mu_B$ is a major challenge both experimentally and theoretically. At vanishing chemical potential, first principle lattice simulations unambiguously demonstrate a smooth crossover from a mostly confined to a mostly deconfined phase, accompanied by a restoration of chiral symmetry \cite{Aoki:2006br,Bazavov:2011nk}. The crossover region sharpens for increasing quark masses and turns in a second order phase transition for critical values of the quarks masses, above which the transition is first order. The same is expected to happen with the chiral transition in the opposite limit of decreasing quark masses: the transition turns first order below a critical value of the quark masses. Although not firmly established by lattice calculations yet \cite{DElia:2018fjp}, this is the expected behavior of a theory with at least three light quark flavors. The situation with two light quarks is more subtle due to the possible role of the axial anomaly~\cite{Pisarski:1983ms}. 

The situation is even less clear at  $\mu_B\neq0$ in the low quark mass region (including the physical point), where standard Monte Carlo algorithms are plagued by the infamous sign problem \cite{deForcrand:2010ys}. 
The typical expectation is that of a line of first order chiral transition at low temperatures ending at a critical point \cite{Stephanov:2004wx}. Firmly establishing the existence of the latter and studying its possible experimental signatures has been the topic of intense theoretical work \cite{Mohanty:2005mv,Stephanov:2008qz,Hatta:2002sj,Roberts:2000aa,Fischer:2018sdj,Fukushima:2013rx,Ding:2015ona} and is among the major physics goals of various present and upcoming experiments \cite{Mohanty:2011nm,Senger:2016wfb,Ablyazimov:2017guv,Sakaguchi:2017ggo}. Methods to circumvent the sign problem on the lattice have been devised but remain, so far, limited to $\mu_B/T\lesssim1$, and no critical endpoint has been firmly established \cite{Fodor:2018wul}. To go beyond, a fruitful proposal has been to supplement lattice results for the quenched gluon dynamics with explicit quark contributions by means of nonperturbative functional methods \cite{Fischer:2012vc,Fischer:2013eca,Eichmann:2015kfa}. Existing works neglect the mesonic degrees of freedom---although baryons have been included \cite{Eichmann:2015kfa}---and find a critical endpoint at a relatively large $\mu_B/T\gtrsim3$. A complementary approach uses phenomenological, Nambu-Jona-Lasinio (NJL) or quark-meson models with various degrees of sophistication \cite{Jakovac:2003ar,Schaefer:2007pw,Fukushima:2008wg,Herbst:2010rf,Resch:2017vjs}. These typically predict a critical endpoint at relatively large $\mu_B/T$, whose precise location, however, varies significantly from one study to another. 
One common weakness of such approaches is that the employed approximations lack a systematic ordering principle. One typically explores the whole phase diagram with truncations  adjusted against lattice data at $\mu_B=0$, far from the region where a critical point is found. 

In the present article, we study this question using a semi-perturbative approach to the infrared dynamics of QCD based on a simple massive extension of the Faddeev-Popov (FP) Lagrangian in the Landau gauge, known as the Curci-Ferrari (CF) model \cite{Curci:1976bt,Tissier:2010ts}. In this context, the gluon mass term is motivated both by the results of lattice simulations \cite{Bogolubsky:2009dc} and by the necessity to modify the FP Lagrangian in the infrared due to Gribov ambiguities \cite{Serreau:2012cg}. The CF model is the simplest renormalizable deformation of the FP Lagrangian and remains under perturbative control down to the deep infrared: The gluon mass screens the standard perturbative Landau pole  and the (running) gauge coupling remains moderate at all scales \cite{Tissier:2010ts,Reinosa:2017qtf}, as observed in lattice simulations. A series of recent studies has shown that the perturbative Curci-Ferrari model gives an accurate description of the phase structure of pure Yang-Mills theories and of QCD with heavy quarks \cite{Reinosa:2014ooa,Reinosa:2015oua,Maelger:2017amh}. The case of light quarks is more delicate because, unlike the couplings in the pure gauge sector, the quark-gluon coupling becomes significant in the infrared \cite{Skullerud:2003qu}. A systematic approximation scheme, nonperturbative in the quark-gluon vertex, has been proposed  in Ref.~\cite{Pelaez:2017bhh}, based on a double expansion in powers of the pure gauge coupling and of the inverse number of colors $1/N_c$. At leading order, this leads to the well-known rainbow equation and, therefore, successfully describes the dynamics of chiral symmetry breaking in the vacuum. There are however too notable differences with respect to the usual implementations of the rainbow equation: first, the gluon exchange is described in terms of a tree-level Curci-Ferrari propagator, and second, higher order corrections are controlled by small parameters, which allows in particular to include renormalization group effects in a consistent way \cite{Pelaez:2017bhh}. 

The purpose of the present work is to extend this approach to nonzero temperature and nonzero chemical potential, paving the way for a systematic study of the predictions of the CF model for the QCD phase diagram.

\section{The rainbow equation in the Curci-Ferrari model}
For simplicity, we study a theory with $N_f$ degenerate quark flavors. At nonzero temperature $T$ and quark chemical potential $\mu=\mu_B/3$, the rainbow equation for the (Euclidean) quark propagator $S$ reads 
\beq
S^{-1}(P) & = & M_0-(i\hat\omega_p-\mu)\gamma_0-i\vec p\cdot \vec \gamma\nonumber\\
\label{eq:rainbow}
& +&\,g^2_0\int_{\hat{Q}}^T\gamma_\mu S(Q)\gamma_\nu\,G_{\mu\nu}(K)\,,
\eeq
where $M_0$ and $g_0$ denote the bare quark mass and quark-gluon coupling, respectively. We have introduced Euclidean momenta $P= (\hat\omega_p,\vec p)$, $Q=(\hat\omega_q,\vec q)$ and $K\equiv P-Q=(\omega_k,\vec{k})$, with $\smash{\omega_n=2\pi nT}$ and $\smash{\hat\omega_n=2\pi(n+1/2)T}$ ($n\in\mathds{Z}$) the bosonic and fermionic Matsubara frequencies respectively. Correspondingly, the bosonic and fermionic Matsubara sums will be denoted
\beq
\int_K^Tf(K)\equiv T\sum_{k\in\mathds{Z}}\int\frac{d^3k}{(2\pi)^3}\,f(\omega_k,\vec k)\,,\\
\int_{\hat Q}^Tf(Q)\equiv T\sum_{q\in\mathds{Z}}\int\frac{d^3q}{(2\pi)^3}\,f(\hat\omega_q,\vec q)\,,
\eeq 
where it is implicitly understood that integrals over the norm of three-dimensional momenta are cut off at a scale $\Lambda$. The matrices $\gamma_\mu$ stand for the Euclidean Dirac matrices, with $\{\gamma_\mu,\gamma_\nu\}=2\delta_{\mu\nu}$, which we choose in the Weyl basis, such that $\gamma_{0,2}^*=\gamma_{0,2}^t=\gamma_{0,2}$ and $\gamma_{1,3}^*=\gamma_{1,3}^t=-\gamma_{1,3}$. Finally, the tree-level gluon propagator is 
\beq
G_{\mu\nu}(K)=\frac{P^\perp_{\mu\nu}(K)}{K^2+m^2}\,,
\eeq 
\vglue-0.2mm
\noindent{with $\smash{P^\perp_{\mu\nu}(K)= \delta_{\mu\nu}-K_\mu K_\mu/K^2}$ the transverse projector and $m$ the Curci-Ferrari mass.}

In Appendix \ref{app:symmetries}, see also \cite{Roberts:2000aa}, we recall that the quark propagator decomposes as (we keep the $\mu$-dependence explicit)
\beq
S(\hat\omega_p,\vec p\,; \mu)=\tilde B+(i\hat\omega_p-\mu)\gamma_0 \tilde A_0+i\vec p\cdot \vec \gamma \tilde A_v+i\gamma_0\vec p\cdot \vec \gamma \,\tilde C,\nonumber\\
\eeq
with any of the components $\smash{\tilde X=\tilde A_0,\tilde A_v,\tilde B}$ or $\tilde C$ depending on $\vec{p}$ only through its norm $p\equiv|\vec{p}|$ and such that
\beq
\tilde X(-\hat\omega_p, p\,; -\mu)=\tilde X(\hat\omega_p, p\,; \mu)\,,\label{eq:C2}\\
\tilde X(-\hat\omega_p, p\,; \mu^*)^*=\tilde X(\hat\omega_p, p\,; \mu)\,.\label{eq:K2}
\eeq
These considerations apply also to the inverse propagator $S^{-1}(\hat\omega_p,\vec{p}\,;\mu)$ which we parametrize as
\beq\label{eq:decomp}
 S^{-1}(\hat\omega_p,\vec{p}\,;\mu)\!=\!B\!-\!(i\hat\omega_p-\mu) \gamma_0 A_0\!-\!i\vec p\cdot \vec \gamma A_v\!-\!i\gamma_0\vec p\cdot \vec \gamma \,C.\nonumber\\
\eeq
We have $X=\Delta\tilde X$, with
\beq
\Delta=B^2+(\hat\omega_p+i\mu)^2A_0^2+p^2(A_v^2-C^2)\,.
\eeq
Projecting Eq.~(\ref{eq:rainbow}) onto the various tensor components, one arrives at a non-linear system of integral equations:
\begin{widetext}
\beq
B(P)&=& M_0 + 3\, g_0^2\,C_F \int_{\hat Q}^T\,\frac{B(Q) }{\Delta(Q)}\frac{1}{K^2+m^2}\,, \label{RforB}
\\
\hat A_0(P)&=& \hat\omega_p+i\mu +  g_0^2\,C_F \int_{\hat Q}^T\,\frac{1}{\Delta(Q)}\frac{1}{K^2+m^2}\Bigg\{ \hat A_{0}(Q) \bigg(1+2\frac{\omega_k^2}{K^2}\bigg) +2\hat A_v(Q) \frac{\omega_k}{K^2} \vec k\cdot\hat{q} \Bigg\},\\
\hat A_{v}(P) &=& p+ g_0^2\,C_F \int_{\hat Q}^T\,\frac{1}{\Delta(Q)}\frac{1}{K^2+m^2}\Bigg\{2 \hat A_{0}(Q) \frac{\omega_k}{K^2}\vec k\cdot\hat{p}+\hat A_v(Q)\, \bigg(\hat{p}\cdot\hat{q} + 2 \frac{(\hat{p}\cdot\vec k)(\vec k\cdot \hat{q})}{K^2}\bigg) \Bigg\},\\
\hat C(P)&=& g_0^2\,C_F \int_{\hat Q}^T\,\frac{\hat C(Q)}{\Delta(Q)}\frac{1}{K^2+m^2}\Bigg\{\hat{p}\cdot\hat{q} \bigg(1-2\frac{\omega_k^2}{K^2}\bigg) -2\frac{(\hat{p}\cdot\vec k)(\vec k\cdot\hat{q})}{K^2} \Bigg\},\label{RforC}
\eeq
\end{widetext}
where $\smash{\hat A_0(P)\equiv (\hat\omega_p+i\mu)A_0(P)}$, $\smash{\hat A_v(P)\equiv pA_v(P)}$ and $\smash{\hat C(P)\equiv p\,C(P)}$, as well as $\smash{\hat p\equiv \vec{p}/p}$. We have also introduced the quadratic Casimir in the fundamental representation $\smash{C_F=4/3}$.
 
In the chiral limit, corresponding to $M_0\to 0$, an unbroken chiral symmetry implies $\smash{B=C=0}$, which obviously solve the (homogeneous) equations (\ref{RforB}) and (\ref{RforC}). This also means that a solution with either $B\neq 0$ or $C\neq 0$ signals the spontanous breaking of chiral symmetry. In what follows, we use $B$ as our order parameter for chiral symmetry breaking since $C=0$ remains an allowed solution (which we stick to) also away from the chiral limit. In order to keep the discussion as simple as possible, we also set the other functions to their tree-level values, $A_0=1$, $A_v=1$.

With this ansatz, the rainbow equation for the quark mass function $B$, Eq. (\ref{RforB}), reads
\beq\label{RB1}
B(P)=M_0+4g^2_0\int_{\hat Q}^T\frac{B(Q)}{Q^2_{i\mu}+B^2(Q)}\frac{1}{(P-Q)^2+m^2}\,,\nonumber\\
\eeq
where we have defined $\smash{Q_{i\mu}\equiv (\hat\omega_q+i\mu,\vec q)}$ and we recall that, in the case of a real chemical potential,
\beq\label{eq:Bsym}
B^*(\hat\omega_p,p;\mu)=B(-\hat\omega_p,p;\mu)=B(\hat\omega_p,p;-\mu)\,,
\eeq
as follows from Eqs.~(\ref{eq:C2}) and (\ref{eq:K2}). In particular, $B$ is real for $\mu=0$ but becomes a priori complex when $\mu$ is non-zero.\footnote{An interesting exception that we shall exploit below is the zero-temperature limit for fixed integer $p$ in $\hat\omega_p$. In this limit, $\hat\omega_p\to 0$ and $B$ becomes real. We mention that it is also real in the case of an imaginary chemical potential.}

\section{Localization}
In this article, we analyse the solutions to Eq.~(\ref{RB1}) by using an approximation scheme called localization \cite{Reinosa:2011cs,Marko:2015gpa}, which we now recall and extend to the problem at hand.

In a given model, the value of the mass function at a given momentum depends on the value of the mass function at any other momentum. However, there can be cases where, in some range of parameters and to a reasonable level of accuracy, the value of the mass function at a particular scale decouples from the rest and obeys, therefore, a simpler, ``localized'' equation. For instance, in Ref.~\cite{Reinosa:2011cs},  the behavior of the mass function at zero momentum was essentially controlled by the zero-momentum mass itself. The self-consistent equation for this zero-momentum mode could be obtained by expanding the mass function about this zero-momentum value in the corresponding integrals. 

Even in cases where there is not a clear argument of why a certain scale could decouple, the localized equations often provide a good qualitative guide. In the present case, it correctly captures the phenomenology of chiral symmetry breaking in the vacuum. 

Of course, rainbow equations similar to Eq.~(\ref{RB1}) have been solved before without the need to resort to localization, even at finite temperature. Localization is however a convenient approach that could be used as a first investigation of more intricate settings such as the rainbow equation in the presence of non-trivial gluonic background (accounting for the interplay between chiral and center symmetry), or the corrections beyond the rainbow approximation, within the systematic expansion scheme alluded to in the Introduction. 

One of the goals of this work is to investigate this strategy in a simpler setting before applying it to these more complicated cases. In order to test the robustness of the approach, we shall consider two types of localizations, referred to below as {\it Euclidean} and {\it physical} respectively, and which we now define more precisely.

\subsection{Euclidean localization}\label{case34}
At finite temperature, it is important to stress that the Euclidean mass function $B$ is defined a priori on the fermionic Matsubara frequencies, which never vanish. We shall therefore localize the mass function at the smallest momentum available, that is, $\smash{B(\hat\omega_1)\equiv B(\hat\omega_1,0;\mu)}$. Since $\smash{B(\hat\omega_1)=B(-\hat\omega_1)^*}$, it is natural -- and in fact crucial, as we illustrate below -- to localize $B(\hat\omega_1)$ together with $B(-\hat\omega_1)$. This means that we should consider two regions in the integrals, the one where it makes sense to expand the mass function about $\hat\omega_1$ and the one where it makes sense to expand it about $-\hat\omega_1$. Therefore, we approximate (\ref{RB1}), with $P=(\hat\omega_1,\vec{0})$ or $P=(-\hat\omega_1,\vec{0})$ by 
\beq
B(\hat\omega_1)&=&M_0+2g_0^2 \int_{\hat Q}^T \,\frac{1}{(\hat\omega_1-\hat\omega_q)^2+q^2+m^2}
\nonumber\\
&&\times \Bigg[\frac{B(\hat\omega_1)}{(\hat\omega_q+i\mu)^2+q^2+B(\hat\omega_1)^2}
\nonumber\\
&&\quad\quad+\frac{B(-\hat\omega_1)}{(\hat\omega_q+i\mu)^2+q^2+B(-\hat\omega_1)^2}\Bigg],\label{rucki}
\eeq
and
\beq
B(-\hat\omega_1)&=&M_0+2g_0^2 \int_{\hat Q}^T \,\frac{1}{(\hat\omega_1+\hat\omega_q)^2+q^2+m^2}
\nonumber\\
&&\times \Bigg[\frac{B(\hat\omega_1)}{(\hat\omega_q+i\mu)^2+q^2+B(\hat\omega_1)^2}
\nonumber\\
&&\quad\quad+\frac{B(-\hat\omega_1)}{(\hat\omega_q+i\mu)^2+q^2+B(-\hat\omega_1)^2}\Bigg],
\eeq
which are easily checked to be compatible with $B(\hat\omega_1)=B(-\hat\omega_1)^*$. It is convenient to work with the real quantities
\beq
B_r & \equiv & \frac{B(\hat\omega_1)+B(-\hat\omega_1)}{2}\,,\\
B_i & \equiv & \frac{B(\hat\omega_1)-B(-\hat\omega_1)}{2i}\,,.
\eeq
In terms of $B_r$ and $B_i$, the rainbow equations for $B(\pm\hat\omega_1)$ read
\beq
B_r &=& M_0 + 2g_0^2 \int_Q^T  \frac{1}{(\hat\omega_1-\hat\omega_q)^2+q^2+m^2} 
\nonumber\\ 
&&\times {\rm Re}\Bigg[ \frac{B_r+iB_i}{(\hat\omega_q+i\mu)^2+q^2+(B_r+iB_i)^2}
\nonumber\\
&&\hspace{0.7cm}+\,\frac{B_r-iB_i}{(\hat\omega_q+i\mu)^2+q^2+(B_r-iB_i)^2}\Bigg], \label{combo}
\eeq
and
\beq
B_i &=& 2g_0^2 \int_Q^T  \frac{1}{(\hat\omega_1-\hat\omega_q)^2+q^2+m^2} 
\nonumber\\ 
&&\times\,{\rm Im}\Bigg[ \frac{B_r+iB_i}{(\hat\omega_q+i\mu)^2+q^2+(B_r+iB_i)^2}
\nonumber\\
&& \hspace{0.7cm}+\,\frac{B_r-iB_i}{(\hat\omega_q+i\mu)^2+q^2+(B_r-iB_i)^2}\Bigg]. \label{combo2}
\eeq
After performing the Matsubara sums and the angular integrals, Eqs.~(\ref{combo})-(\ref{combo2}) rewrite in the simple form 
\beq
M_0 &=& B_r-\frac{g_0^2}{2\pi^2}\,{\rm Re}\Big[(B_r+iB_i)F\big(B_r+iB_i\big)\nonumber\\
& & \hspace{2.0cm}+\,(B_r-iB_i)F\big(B_r-iB_i\big)\Big],\\
0 &=& B_i-\frac{g_0^2}{2\pi^2}\,{\rm Im}\Big[ (B_r+iB_i)F\big(B_r+iB_i\big)\nonumber\\
& & \hspace{2.0cm}+\,(B_r-iB_i)F\big(B_r-iB_i\big)\Big],
\eeq
where, for notational convenience, we have defined 
\beq\label{eq:FB}
F(B)\equiv F_{\rm vac}(B)+F_{\rm th}(B)\,,
\eeq 
with
\beq
F_{\rm vac}(B) &\equiv& \int_0^\Lambda{\rm d}q\,\frac{q^2}{\varepsilon_q^m\varepsilon_q^B}\frac{1}{\varepsilon_q^B+\varepsilon_q^m}\,,\\
F_{\rm th}(B)&\equiv&\int_0^\Lambda{\rm d}q\,\frac{q^2}{\varepsilon_q^m\varepsilon_q^B}\Bigg\{\frac{(\mu-i\hat\omega_1)^2/(\varepsilon_q^B+\varepsilon_q^m)}{(\varepsilon_q^B+\varepsilon_q^m)^2-(\mu-i\hat\omega_1)^2} \nonumber\\
& & \hspace{2.0cm}+\,\frac{\varepsilon_q^B n^{(-)}_{ \varepsilon_q^m}}{ (\varepsilon_q^B)^2-( \varepsilon_q^m-\mu+i\hat\omega_1)^2}\nonumber\\
& & \hspace{2.0cm}+\,\frac{\varepsilon_q^B n^{(-)}_{ \varepsilon_q^m}}{(\varepsilon_q^B)^2-( \varepsilon_q^m+\mu-i\hat\omega_1)^2}\nonumber\\
& & \hspace{2.0cm}-\,\frac{\varepsilon_q^m n^{(+)}_{ \varepsilon_q^B-\mu}}{ (\varepsilon_q^m)^2-( \varepsilon_q^B-\mu+i\hat\omega_1)^2}\nonumber\\
& & \hspace{2.0cm}-\,\frac{\varepsilon_q^m n^{(+)}_{ \varepsilon_q^B+\mu}}{(\varepsilon_q^m)^2-( \varepsilon_q^B+\mu-i\hat\omega_1)^2}\Bigg\},\label{Fomega}\nonumber\\
\eeq
with $\varepsilon^{x}_{y}\equiv \sqrt{x^2+y^2}$ and where $n^{(\pm)}_x=(e^x\pm1)^{-1}$ denote the Bose-Einstein and Fermi-Dirac distributions. We note that $F(B^*)$ is not the complex conjugate of $F(B)$, due to the dependence on $i\hat\omega_1$ which we leave, however, implicit in what follows.

\subsection{Physical localization}
One inconvenient aspect of the Euclidean localization is that one has to deal with two variables, $B_r$ and $B_i$. This prevents the definition of a potential associated to the localized equations, due to the fact that the latter do not comply with the Cauchy-Riemann conditions. Although many features of the phase diagram do not require the existence of an underlying potential, it is convenient to find a setting where one deals with only one variable instead of two.\footnote{We will see below that, after renormalization, there is one particular way to achieve this within the Euclidean localization. There are also certain limits, such as $T\to 0$ or $\mu\to 0$, where this becomes possible due to the fact that $B_i\to 0$.}  Another drawback of the Euclidean localization is that it involves the first Matsubara frequency $\pi T$. Therefore, we expect its quality to decrease as the temperature is increased.

One way to try to cope with these limitations is to consider a localization based on the retarded mass function
\beq\label{eq:ret_prescription}
B_R(p_0,p)=B(\hat\omega_p\to-i(p_0+\mu)+0^+,p)\,,
\eeq
evaluated for $\smash{p_0=0}$ and $\smash{p=0}$. The reason for the presence of $\mu$ in the prescription to obtain the physical retarded Green's function is recalled in Appendix \ref{app:mu_prescription}, see also Ref.~\cite{Laine:2016hma}. We find that, for $B$ real and smaller than $m$, the corresponding analytic continuation of $F(B)$ in Eq.~(\ref{eq:FB}) is real in the limit $p_0\to 0$.\footnote{One should pay attention to the fact that what is really continued are not $B_r$ and $B_i$, but rather $B(\omega)$.} Therefore, the equation for $B_i$ becomes compatible with the solution $B_i=0$, which we assume from now on, and the equation for $B_r\equiv B$ reduces down to
\beq\label{eq:RBbare}
M_0=\left(1-\frac{g^2_0}{{\pi^2}}\Big[F_{\rm vac}(B)+F_{\rm th}(B)\Big]\right)\!B\,,
\eeq
where $F(B)$ is to be evaluated with $\hat\omega_1\to 0$ and $\mu\to 0$ in the energy denominators. While $F_{\rm vac}(B)$ remains the same as above, we now have
\be
F_{\rm th}(B) = \frac{2}{B^2-m^2}\!\int_0^\infty \!dq\,q^2\!\left(\frac{n^{(-)}_{\varepsilon^{m}_{q}}}{\varepsilon^{m}_{q}}+\frac{n^{(+)}_{\varepsilon^{B}_{q}-\mu}+n^{(+)}_{\varepsilon^{B}_{q}+\mu}}{{2\varepsilon^{B}_{q}}}\right)\!.
\ee
One price to pay for this choice of localization is the singularity at $B=m$, which is regulated in the retarded self-energy through $B^2-m^2\to B^2-m^2+i0^+$. We ignore this issue and restrict our analysis to cases where $B<m$. This is justified both because the physically relevant values in the vacuum fall in this range (see below) and because most of our discussions below concern the vicinity of the symmetric point $B=0$.\footnote{On the other hand, the constraint $B<m$ does not allow for a smooth continuation from the chiral limit to the heavy quark limit since $B$ is always constrained to be less than $m$. This is one of the reasons why we also develop the Euclidean localization.} We note finally  that, interestingly (although not surprisingly), the localized equation \eqn{eq:RBbare} reduces to the corresponding mean-field gap equation of a NJL-type model with an effective nonlocal four-fermion vertex corresponding to a massive gluon exchange.

\section{Results}
We now investigate our predictions for the phase diagram using the two types of localizations, starting with the physical localization.

\subsection{Physical localization}
Consider first the chiral limit, corresponding to $\smash{M_0\to 0}$. For large enough values of the coupling, \Eqn{eq:RBbare} admits nontrivial, symmetry breaking solutions on top of the chirally symmetric solution $B=0$.  It is convenient to parametrize the equation in terms of the dynamical mass in the vacuum, $B_0\equiv B(T=0,\mu=0)$, given by
\beq\label{eq:g0B0}
F_{\rm vac}(B_0)=\pi^2/g_0^2\,,
\eeq
where we see that symmetry-breaking solutions exist only if $g_0^2>{\pi^2}/F_{\rm vac}(0)$.
We use \Eqn{eq:g0B0} to trade the bare quark-gluon coupling $g_0$ for the (ultraviolet finite) quark mass $B_0$.
The rainbow equation rewrites as
\be\label{RBe}
0=B\left[F_{\rm vac}(B_0)-F_{\rm vac}(B)-F_{\rm th}(B)\right]\equiv 2BR(B^2)\,,
\ee
where we note that the cut-off can now be sent to infinity:
\be\label{Fvace}
F_{\rm vac}(B_0)-F_{\rm vac}(B)=\frac{B^2\ln(B/m)}{{2}(B^2-m^2)}-\frac{B_0^2\ln (B_0/m)}{{2}(B_0^2-m^2)}.
\ee

\begin{figure}[t]
	\centering
	\includegraphics[width=0.42\textwidth]{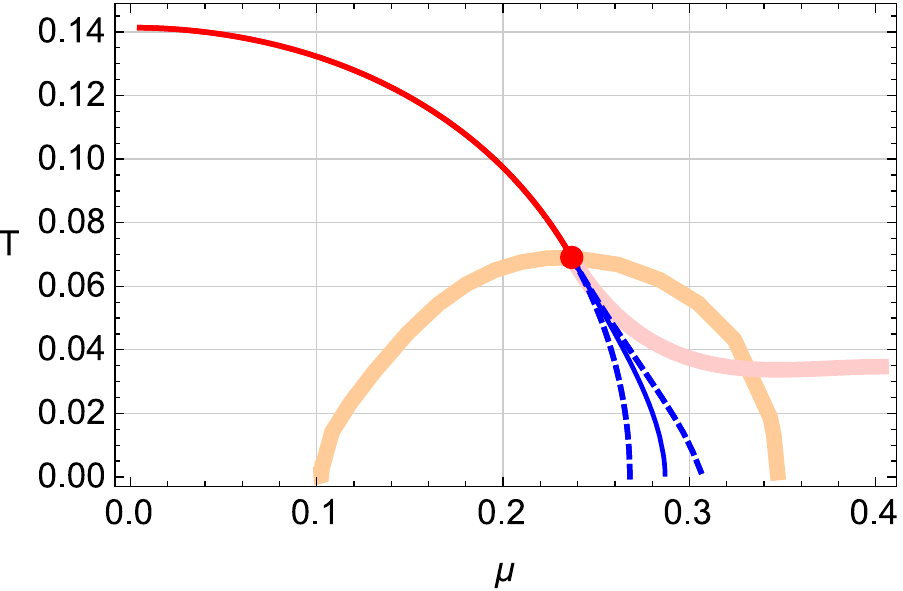}
	\caption{Phase diagram in the chiral limit (all scales in GeV), as obtained from the physical localization of the rainbow equation, with $B_0=0.3$~GeV and $m=0.5$~GeV. The line of second-order transitions (solid red) turns into a line of first-order transitions (solid blue) at the tricritical point (red dot). The dashed curves are the corresponding spinodal lines. The orange line shows the location of the tricritical point as a function of the gluon mass (as described in the main text). Away from the chiral limit, the tricritical point turns into a critical endpoint, whose position follows the pink line as the bare quark mass is increased.}
	\label{fig:tric}
\end{figure}

It is useful to interpret \Eqn{RBe} as deriving from a chirally symmetric potential $W(B^2)$, with $\partial_B W(B^2)=2BR(B^2)$, that is $W'(B^2)=R(B^2)$. The absolute minima of $W(B^2)$ then determine the  state of the system. At $\mu=0$, one easily checks that the nontrivial minimum, equal to $B_0$ in the vacuum, decreases with increasing temperature and continuously reaches $B=0$ at a critical temperature. This extends in a critical line of second-order phase transitions $T_c(\mu)$ in the $(\mu,T) $ plane, defined by the condition $W'(0)=0$, as shown in \Fig{fig:tric}. Depending on the parameters, this critical line can turn into a line of first-order transitions at a tricritical point, defined by the conditions $\partial_B^{{(2n)}} W(B^2)|_{B=0}=0$ for {$n=1,2$} or, equivalently,
\beq\label{tric}
W'(0)=W''(0)=0\,.
\eeq
The first-order line is then determined from 
\beq\label{eq:first}
W'(B_{\rm min}^2)=W(B_{\rm min}^2)-W(0)=0\,,
\eeq
where $B_{\rm min}$ is the nontrivial minimum at the transition. The associated lower and upper spinodals are respectively defined by
\beq\label{spinodal}
W'(0)=0\quad{\rm and}\quad W'(B_{\rm sp}^2)=W''(B_{\rm sp}^2)=0\,,
\eeq
with $B_{\rm sp}$ the location of the nontrivial metastable state at the upper spinodal.
The two spinodals flank the first order line and merge at the tricritical point, beyond which the lower spinodal becomes the critical line.
The equation governing the critical and lower spinodal lines is easily obtained as
\be\label{eq:critical}
\mu^2(T)=m^2\frac{B_0^2\ln (B_0/m)}{B_0^2-m^2}-4\int_0^\infty \!\!dq\,q^2\frac{n^{(-)}_{\varepsilon^{m}_{q}}}{\varepsilon^{m}_{q}}-\frac{\pi^2}{3}T^2\,,
\ee
which is a strictly concave line. For low enough temperatures $T/m\ll1$, the gluonic thermal contribution (second term) in \Eqn{eq:critical} is negligible and one obtains the approximate expression
\be \label{eq:criticalapprox}
\mu^2(T)\approx m^2\frac{B_0^2\ln (B_0/m)}{B_0^2-m^2}-\frac{\pi^2}{3}T^2\,.
\ee
This is similar to the result obtained in the quark-meson model with a large-$N_f$ approximation~\cite{Jakovac:2003ar}.

We note that there is an ambiguity in the definition of the potential $ W(B^2)$ since neither the solutions of \Eqn{RBe} nor their convexity are altered by the replacement  
$ W'(B^2)\to f_+(B^2)R(B^2)$ with $f_+(B^2)$ a differentiable and strictly positive function. 
Interestingly, because the conditions \eqn{tric} and \eqn{spinodal} only involve $W'$ and its derivatives, they are, in fact, independent of the function $f_+$ and so are, thus, the spinodal lines, the line of second-order transition, and, of course, the tricritical point where these lines meet. Only the line of first-order transition explicitly depends on $f_+$ through the second condition in \Eqn{eq:first}. However, it always lies in between the two spinodals.

\begin{figure}[t]
	\centering
	\includegraphics[width=0.45\textwidth]{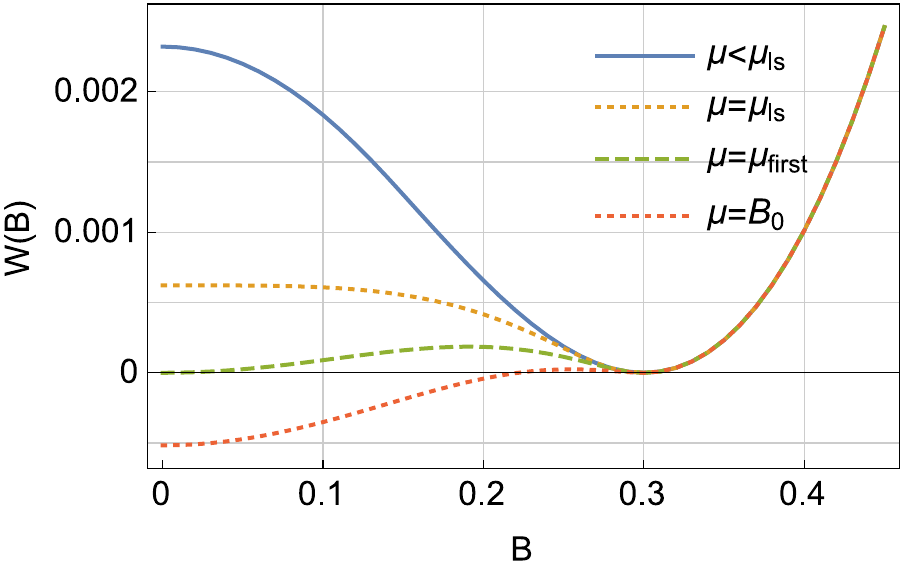}
	\caption{Illustration of the Silver-Blaze property: as long as $\mu$ is below the first order transition, the minimum of the potential sits at $B=B_0=0.3$\,GeV. For $\mu_{\rm first}<\mu<B_0$, there is still an extremum at $B=B_0$, but it is not anymore the absolute minimum. For $\mu>B_0$, this local minimum moves away from $B_0$ and disaspears at the upper spinodal $\mu_{us}$ (slightly above $\mu=B_0$, not shown).}
	\label{fig:SB}
\end{figure}

In Fig.~\ref{fig:tric}, we show our results for the phase diagram in the chiral limit. We use the typical values  $m=500$~MeV and $B_0=300$~MeV, motivated by the study of dynamical chiral symmetry breaking in the Curci-Ferrari model in the vacuum \cite{Pelaez:2017bhh}. We find a tricritical point located at $(\mu,T)\approx (237\,{\rm MeV},69\,{\rm MeV})$. The transition at zero chemical potential occurs at $T_c(\mu=0)\approx 141$~MeV and the two spinodals meet the $T=0$ axis for $\mu\approx268$~MeV, and $\mu\approx 305$~MeV, respectively. This gives an estimate of the first-order transition line at most at the 10\% level, independently of the function $f_+$ (the estimate improves as one approaches the tricritical point). For the choice $f_+=1$, the $T=0$ transition point is at {$\mu\approx 287$~MeV}.  

In fact, \Eqn{RBe} greatly simplifies at $T=0$, where 
\be\label{Fthe}
F_{\rm th}(B) = \frac{B^2/2}{B^2-m^2}\!\left[\frac{\mu}{B}\sqrt{\frac{\mu^2}{B^2}-1}-\cosh^{-1}\left(\frac{\mu}{B}\right)\right]\!,
\ee
if $\mu\ge B$ and $F_{\rm th}(B)=0$ otherwise. In particular, we see that the value of the order parameter below the transition point is independent of $\mu$, $B(T=0,\mu)=B_0$, until it jumps to $B=0$ in the symmetric phase. This is known as the Silver Blaze property \cite{Cohen:2003kd,Marko:2014hea}, which we illustrate in Fig.~\ref{fig:SB}.\footnote{We mention that the Silver Blaze property should in princple extend only up to the the first singularity on the $T=0$ axis, namely the nuclear liquid-gas transition. Here, however, our level of description does not capture the corresponding dynamics, and the Silver Blaze property extends further. We also mention that the $\mu$-independence below the first singularity applies in principle only to $0$-point functions. For higher $n$-point functions, it takes a more general form as shown in \cite{Marko:2014hea}. However, due to the presence of $\mu$ in the retarded prescription (\ref{eq:ret_prescription}), it can be argued that the Silver-Blaze property applies to retarded Green functions as it does for $0$-point functions.} Also, using Eqs.~\eqn{Fvace} and \eqn{Fthe}, we can determine the values of the gluon mass for which there exists a tricritical point. The latter reaches the $T=0$ axis for some values of the ratio $x=m/B_0$. Defining  $u=(\ln x^2)/(x^2-1)$, we get the condition $u=1+\ln(2u)$, which has two solutions in $\mathds{R}^+$,  $u_\pm=u(x_\pm)$, with $x_+x_-=1$. One finds $x_-=\sqrt{u_+/u_-}\approx 0.294$ and $x_+\approx 3.398$. The corresponding values of $y=\mu/B_0$ are given by \Eqn{eq:critical} at $T=0$, $y^2=x^2u/2$, yielding $y_-=\sqrt{u_+/2}\approx0.340$ and $y_+=y_-/x_-\approx1.157$. There exists a tricritical point  iff $x\in[x_-,x_+]$ as shown in \Fig{fig:tric}.\\

\begin{figure}[t]
	\centering
	\includegraphics[width=0.4\textwidth]{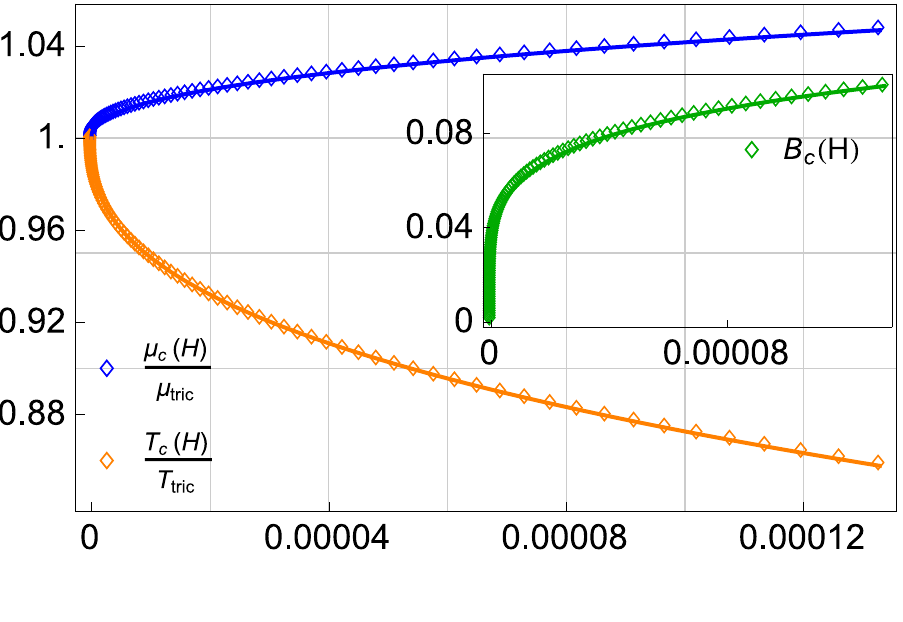} 
	\caption{The approach of the critical quantities $\mu_c(H)$, $T_c(H)$, and $B_c(H)$ to the chiral limit shows mean field tricritical scaling. The solid lines are power law fits of the form $A_c(H)-A_{\rm tric}\propto H^{\omega_{\!A}}$ for $A=\mu,T,B$ with mean field exponents $\omega_T=\omega_\mu=2/5$ and $\omega_B=1/5$.}
	\label{fig:scaling}
\end{figure}

\begin{figure}[t]
	\centering
	\includegraphics[width=0.4\textwidth]{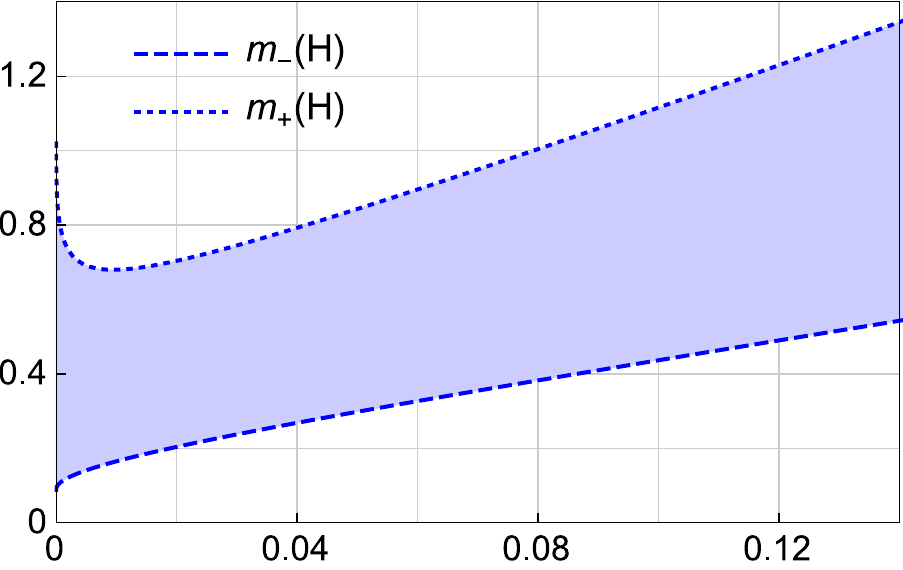}
	\caption{Allowed values of the gluon mass $m$ for a (tri)critical point to exist  as a function of $H$ (blue area). All scales in GeV and $m_\pm\equiv B_0 x_\pm$.}
	\label{fig:mminmax}
\end{figure}

Let us now move away from the chiral limit. Equation~\eqn{RBe} now reads $2 B R(B^2)=H$, where $H\equiv\pi^2 M_0/g_0^2$ needs to be seen as a finite parameter controlling the departure from the chiral limit. For $H\neq0$, the second-order transitions turn into crossovers and the tricritical point becomes a critical endpoint terminating a first-order line. Writing the potential $V(B)=-{H}B+W(B^2)$, the conditions for a critical point are 
\beq\label{crit}
V'(B_c)=V''(B_c)=V'''(B_c)=0\,,
\eeq
from which one extracts $B_c$, $T_c$, and $\mu_c$ for each ${H}$. To this aim, it is convenient to vary $B$, determine  $T_c(B)$ and $\mu_c(B)$ from the last two conditions in \Eqn{crit}---that do not involve ${H}$---and then deduce $H(B)$ from the first condition. Inverting this relation, one then gains access to $T_c({H})$ and $\mu_c({H})$. In particular, we find that the approach to tricriticality is governed by mean field exponents; see Fig.~\ref{fig:scaling}. This is expected because the potential is regular around $B=0$. The trajectory of the critical endpoint in the phase diagram, shown in Fig.~\ref{fig:tric}, exhibits a nonmonotonous behavior of $T_c$ as a function of $\mu$, similar to that observed in Ref.~\cite{Hatta:2002sj} using an approach based on the Cornwall-Jackiw-Tomboulis effective potential. Finally, \Fig{fig:mminmax} shows the interval $[x_-(H),x_+(H)]$ compatible with a critical point for each value of $H$. Interestingly, in the physical localization considered here, the Curci-Ferrari mass should be neither too large nor too small for a critical end point to exist. 

\subsection{Euclidean localization}
Let us now investigate the Euclidean localization, that allows us to test the robustness of the previous features.

To renormalize the equations in this case, we proceed as follows. First, we note that the function $F_{\rm vac}(B)$ possesses a logarithmic UV divergence which however does not depend on $B$. It is then readily checked that the equation for $B_r$ has a divergence proportional to $B_r$. This divergence can be absorbed into a redefinition of the bare coupling as follows. We divide the corresponding equation by $g_0^2$ and set
\beq\label{eq:g0}
\frac{1}{g_0^2}=\frac{1}{g^2}+\frac{1}{\pi^2}F_{\rm vac}(B_\star)\,,
\eeq
where the renormalized coupling $g$ should be interpreted as being defined at the (real) renormalization scale $B_\star$. Introducing $H\equiv \pi^2M_0/g_0^2$ as before, the renormalized equation for $B_r$ takes the form
\beq\label{eq:Br_ren}
H &=& \frac{\pi^2}{g^2} B_r-\frac{1}{2}\,{\rm Re}\Big[(B_r+iB_i)\tilde F\big(B_r+iB_i\big)\nonumber\\
& & \hspace{2.0cm}+\,(B_r-iB_i)\tilde F\big(B_r-iB_i\big)\Big],
\eeq
with $\tilde F(B)=F_{\rm vac}(B)-F_{\rm vac}(B_\star)+F_{\rm th}(B)$.

As far as the equation for $B_i$ is concerned, it is easily checked that it is finite, for any fixed $g_0$. Therefore, using the redefinition (\ref{eq:g0}) or the bare coupling is problematic here since it leads to a spurious cut-off dependence. This problem is rooted in the localization procedure which does not commute with the renormalization procedure, see \cite{Marko:2015gpa} for more details. However, we can always define a renormalized localized scheme by replacing $g_0$ by $g$ in the equation for $B_i$:
\beq\label{eq:Bi_ren}
0 &=& \frac{\pi^2}{g^2} B_i-\frac{1}{2}\,{\rm Im}\Big[ (B_r+iB_i)\tilde F\big(B_r+iB_i\big)\nonumber\\
& & \hspace{2.0cm}+\,(B_r-iB_i)\tilde F\big(B_r-iB_i\big)\Big].
\eeq
We note that, in this equation, one can interchangeably use $F$ or $\tilde F$.

With this choice of renormalization, we can eventually express the equations in terms of the vacuum mass in the chiral limit, $B_0$, see Eq.~(\ref{eq:g0B0}). It is related to $g$ and $B_\star$ by
\beq\label{eq:gofBs}
F_{\rm vac}(B_0)-F_{\rm vac}(B_\star)=\pi^2/g^2\,.
\eeq
Replacing $g^2$ in that form, one checks that the first term in the RHS of (\ref{eq:Br_ren}) disappears while the scale $B_\star$ in $\tilde F(B)$ is replaced by $B_0$. Thus, Eq.~(\ref{eq:Br_ren}) does not depend on $B_\star$. However a scale dependence remains in equation (\ref{eq:Bi_ren}), as expected at a given order of approximation. Below, we shall test the dependence of our results on the renormalization scale $B_\star$.

Let us also mention that, because of the localization procedure, the coupled gap equations (\ref{eq:Br_ren}) and (\ref{eq:Bi_ren}), which we denote formally as $\smash{H={\cal R}_r(B_r,B_i)}$ and $\smash{0={\cal R}_i(B_r,B_i)}$ in what follows, cannot be seen as deriving from a potential, because, in general the Cauchy-Riemann condition $\partial{\cal R}_r/\partial B_i=\partial {\cal R}_i/\partial B_r$ is not satisfied. However, certain features of the phase diagram can be defined without the need of a potential because they correspond to the merging of different solutions of the gap equations. For instance, suppose that we want to investigate whether $B_r$ becomes critical. To this purpose, we solve for $B_i$ as a function of $B_r$ from its gap equation:
\beq\label{eq:eqone}
0={\cal R}_i(B_r,B_i(B_r))\,,
\eeq
and construct a potential for $B_r$ by integrating
\beq\label{eq:eqtwo}
V'(B_r)={\cal R}_r(B_r,B_i(B_r))\,.
\eeq
In the chiral limit, the criticality condition reads $0=V''(0)$. Simple algebra using (\ref{eq:eqone}) and (\ref{eq:eqtwo}) leads to the condition
\beq\label{eq:critico}
0=\left.\frac{\partial{\cal R}_r}{\partial B_r}-\frac{\partial{\cal R}_r}{\partial B_i}\left(\frac{\partial{\cal R}_i}{\partial B_i}\right)^{-1}\frac{\partial {\cal R}_i}{\partial B_r}\right|_{B_r=B_i=0}\,.
\eeq
Now, it is easily verified that $\smash{{\cal R}_r(B_r,B_i)={\cal R}_r(B_r,-B_i)}$, from which it follows that $\smash{\partial {\cal R}_r/\partial B_i|_{B_i=0}=0}$. Similarly, writing $\smash{{\cal R}_i(B_r,B_i)\equiv B_i\pi^2/g^2+\tilde {\cal R}_i(B_r,B_i)}$, we have $\smash{\tilde {\cal R}_i(B_r,-B_i)=\tilde {\cal R}_i(B_r,B_i)}$ and therefore $\smash{\partial {\cal R}_i/\partial B_i|_{B_i=0}=\pi^2/g^2}$. From these remarks, it follows that the condition for a critical point in the chiral limit simplifies to
\beq
0=\left.\frac{\partial{\cal R}_r}{\partial B_r}\right|_{B_r=B_i=0}\,.
\eeq
The same equation defines the lower spinodal in the case of a first order phase transition. The upper spinodal is also determined from (\ref{eq:critico}) but without evaluating it for $B_r=B_i=0$ and coupling it to the gap equation for $B_r$. Finally, the tricritical point is determined from conditions $0=V''(0)=V^{(4)}(0)$. We find
\beq
0 & = & \sum_{u,v,w} \!\!\left.\frac{\partial^3 {\cal R}_r}{\partial B_u \partial B_v \partial B_w}\right|_{B=0} \!\!\!\!\!\!\Delta B_r\,\Delta B_u \,\Delta B_v \,\Delta B_w\,,\label{fouur}
\eeq
where the indices $u$, $v$, and $w$ take the values $r$ or $i$, and $\Delta B=(\partial{\cal R}_i/\partial B_i,-\partial{\cal R}_i/\partial B_r)_{B_r=B_i=0}$. We have again made use of $\partial {\cal R}_r/\partial B_i|_{B_i=0}=0$. This formula simplifies further because $\partial^3 {\cal R}_r/\partial B_r^2\partial B_i|_{B_i=0}=\partial^3 {\cal R}_r/\partial B_i^3|_{B_i=0}=0$. We note however that we are not able to fully eliminate $B_i$, contrary to what happened for the critical point [see below for a particular limit where this becomes possible].\\

Our results in the chiral limit (with $B_\star=1$\,GeV) are summarized in Table \ref{tab:caseschiral} and compared to the results in the physical localization as well as to the results in other approaches. The last column shows the values of $\mu$ at which the lower and upper spinodals ($\mu_{\rm ls}$ and $\mu_{\rm us}$) and the first order transition ($\mu_{\rm first}$) are reached for $T=0$. As already discussed in the previous section, the value of $\mu_{\rm first}$ has the largest uncertainty since it depends on the potential which is not uniquely defined in our approach. Also, it may look surprising that we could obtain a value $\mu_{\rm first}$ in the Euclidean localization case since there is no potential in this case compatible with the gap equations. However, in the $\smash{T\to 0}$ limit, it is readily checked using (\ref{eq:Bsym}) that $B_i$ vanishes. We note also that, along the $T=0$ axis, $B\equiv B_r$ is not constant below the transition. This is of course not in contradiction with the Silver-Blaze property since only $0$-point functions should be constant. The case of the physical localization is a bit peculiar since the retarded prescription (\ref{eq:ret_prescription}) with $\mu$ included makes the retarded function behave like a $0$-point function as far as the Silver-Blaze property is concerned.
\begin{table}[t!]
  \centering
  	\begin{tabular}{|c || c | c || c ||| c | c | c|} 
		\hline
		 Chiral limit ($H=0$)  & $\mu_{\rm tric}$ & $T_{\rm tric}$ & $T_{\rm c}$ & $\mu_{\rm ls}$ & $\mu_{\rm first}$ & $\mu_{\rm us}$ \\ [0.5ex] 
		\hline\hline
		 Physical localization & 237  & 69 & 141 & 268 & 287 & 305 \\
		\hline
	         Euclidean localization & 318  & 64  & 150 & 346 & 365  & 376  \\
		\hline
	         Jakovac {\it et al.} \cite{Jakovac:2003ar} & 280  & 60  & 140\\
		\cline{1-4}
	         Schaefer {\it et al.} \cite{Schaefer:2004en} & 251 & 52 & 142  \\
		\cline{1-4}
	         Hatta {\it et al.} \cite{Hatta:2002sj} & 209 & 107 &  -- \\
		\cline{1-4}
		Qin {\it et al.} \cite{Qin:2010nq} A & 140 & 110 & 124  \\
		\cline{1-4}
		Qin {\it et al.} \cite{Qin:2010nq} B & 130 & 120 & 133  \\
		\cline{1-4}
		Costa {\it et al.} \cite{Costa:2008yh}& 286 & 112 & 215\\
		\cline{1-4}
	\end{tabular}
	\caption{Results in the chiral limit for the two considered localizations, in comparison to benchmark literature findings.  All values are given in MeV. For the Euclidean localization, we have chosen a renormalization scale $B_\star=1$\,GeV.}
	\label{tab:caseschiral}
\end{table}

\begin{figure}[t]
	\centering
	\includegraphics[width=0.23\textwidth]{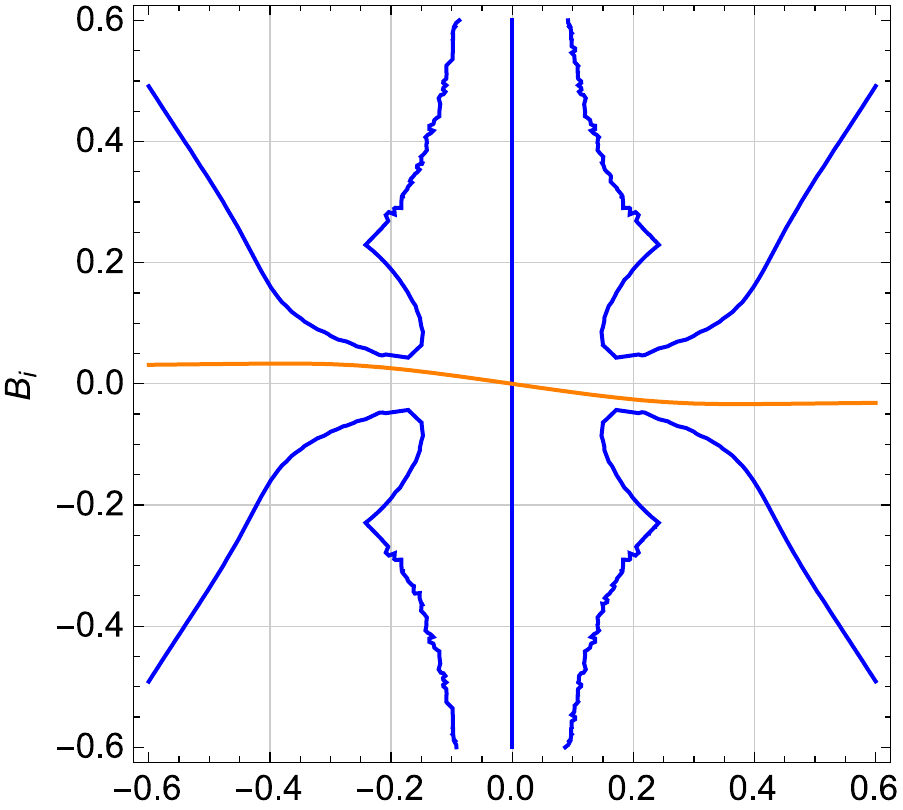}\quad
	\includegraphics[width=0.214\textwidth]{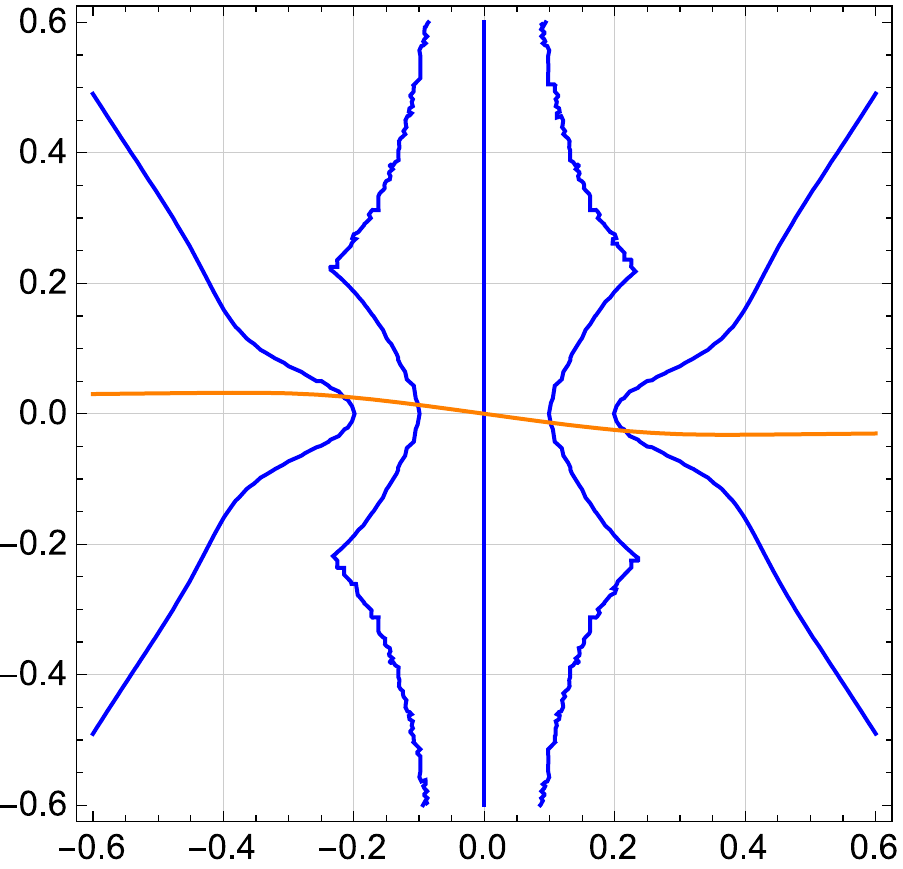}\\
	\vglue4mm
	\includegraphics[width=0.23\textwidth]{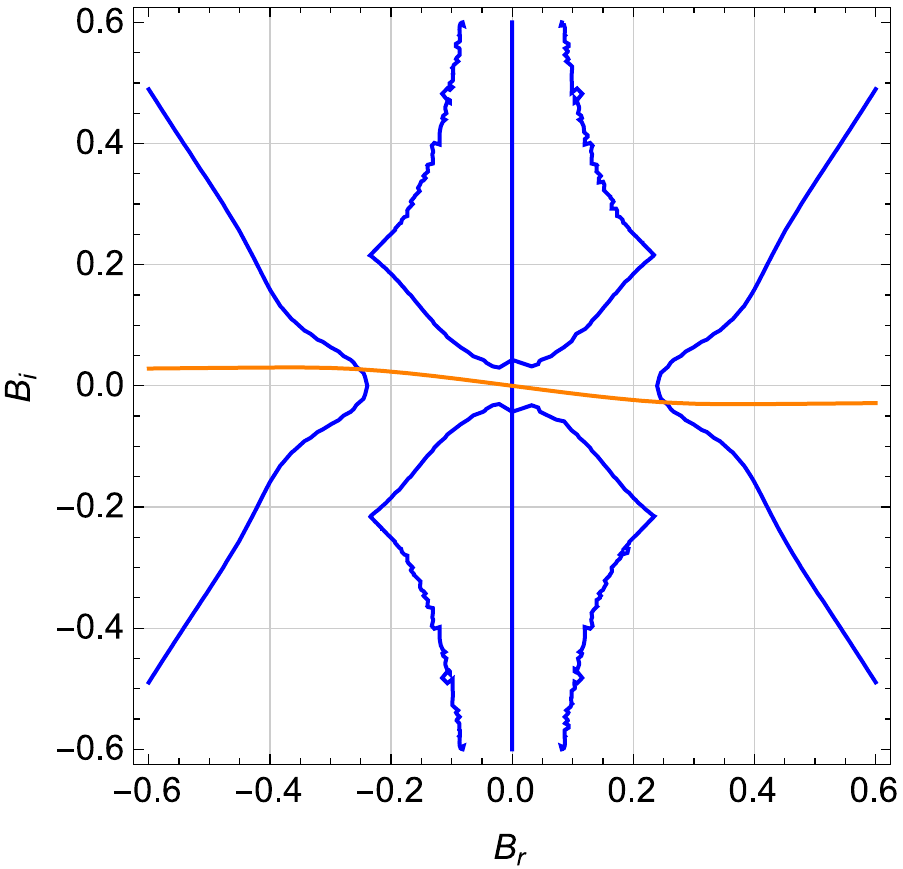}\quad
	\includegraphics[width=0.214\textwidth]{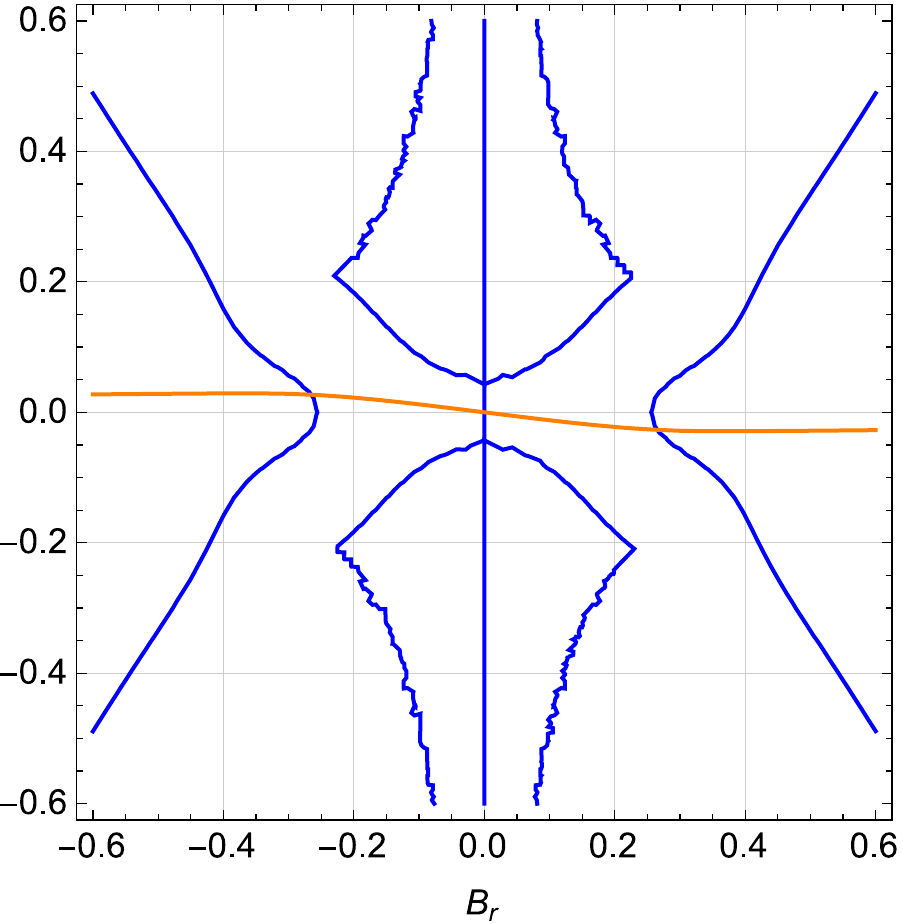}
	\caption{$0$-level plots of the gap equations in the chiral limit in the plane $(B_r,B_i)$ for decreasing temperatures, with $m=0.5$, $B_0=0.3$ and $B_\star=1$ (all units in GeV). We have also chosen $\mu=0.33$ and $T$ takes the values $0.055$, $0.053$, $0.050$ and $0.048$. The wiggling in the curves is due to the presence of singularities of the gap equations in the plane $(B_r,B_i)$ that lead to additional zeros of ${\cal R}_r$ or ${\cal R}_i$ (not visible). Fortunately, the actual solutions of the gap equations, corresponding to simultanous zeros of ${\cal R}_r$ and ${\cal R}_i$, are located far from these regions.} 
	\label{fig:contours}
\end{figure}

We mentioned above that it was crucial to localize simultanously in $B(\hat\omega_1)$ and $B(-\hat\omega_1)$. Let us illustrate this point here. In Fig.~\ref{fig:contours}, we show the curves $\smash{{\cal R}_r(B_r,B_i)=0}$ and $\smash{{\cal R}_i(B_r,B_i)=0}$ for decreasing temperatures and a large enough chemical potential. The crossings correspond to the various possible solutions in the chiral limit and, because the chemical potential is chosen large enough, we should observe a first order transition pattern. Let us now see how this comes about. The first plot is at a temperature right above the upper spinodal, that is the appearance of a new crossing between the curves $\smash{{\cal R}_r(B_r,B_i)=0}$ and $\smash{{\cal R}_i(B_r,B_i)=0}$, at which two new extrema are about to appear.\footnote{In fact, there are four such extrema since, in the chiral limit, the problem is symmetric under $(B_r,B_i)\to(-B_r,-B_i)$. But it is then enough to restrict to $B_r>0$.} Below this temperature, the various branches making the curve $\smash{{\cal R}_r(B_r,B_i)=0}$ fuse and reorganize, in such a way that, at an even lower temperature, a second spinodal occurs at $B=0$ where two extrema merge. We observe that the proper realization of this scenario requires not only the various branches to fuse at some temperatures, but also that the number of intersections of the curves $\smash{{\cal R}_r(B_r,B_i)=0}$ and $\smash{{\cal R}_i(B_r,B_i)=0}$ changes from one to five, and then to three, as the temperature is decreased. Had we performed the localization only with respect to $B(\hat\omega_1)$, that is by writing only the first term in Eq.~(\ref{rucki}) and taking the real and imaginary part of the corresponding equation, this second requirement would not be fulfilled, as we have checked explicitly.\\

Regarding the renormalization scale dependence of our results, we observe numerically that the critical/lower spinodal line does not depend on the scale $B_\star$. This is no surprise since the corresponding equation (\ref{eq:critico}) depends only on ${\cal R}_r(B_r,0)$ which, has we have already argued above is $B_\star$ independent. In contrast, the position of the tricritical point along (\ref{eq:critico}), or even the other spinodal emerging from this point do depend on $B_\star$. We note, however, that the inverse coupling $1/g^2$ diverges positively as the renormalization scale is taken to infinity, see Eq.~(\ref{eq:gofBs}). Since there is no other dependence with respect to $B_\star$ in Eq.~(\ref{eq:Bi_ren}), it follows that $B_i$ should approach $0$ in this limit and all relevant features (boundary lines, tri/critical points, \dots) should converge to a certain limit, obtained by considering a single gap equation (\ref{eq:Br_ren}) in which one sets $B_i=0$ from the start.\footnote{This equation does not become trivial in the limit $\smash{B_\star\to\infty}$ because, in this case, the $B_\star$-dependence of $1/g^2$ is cancelled by the corresponding $B_\star$-dependence hidden in $\tilde F$. We also mention that the equation obtained in the $B_\star\to\infty$ limit is nothing but the one we would have obtained by applying the na\"ive renormalization and sending the cut-off to infinity. Indeed the remaining cut-off dependence in the equation for $B_i$, only present in the term containing $1/g_0^2$, would enforce $B_i\to 0$ as $\Lambda\to\infty$.} Take for instance the tricritical point. Because $\Delta B_r=-\pi^2/g^2\gg \Delta B_i$ in the limit $B_\star\to\infty$, Eq.~(\ref{fouur}) becomes
\beq
0=\left.\frac{\partial^3 {\cal R}_r}{\partial B_r^3}\right|_{B=0}\,,
\eeq
which is indeed the condition for a tricritical point if one restricts from the beginning to Eq.~(\ref{eq:Br_ren}) with $B_i=0$.

The relative difference between the tricritical values for $B_\star=1$\,GeV and $B_\star\to\infty$ are found to be about a few percent. This indicates a controlled renormalization scale dependence and allows us from now on to work in a simplified picture in the $B_\star\to\infty$ limit. In particular, in this limit, we can associate a potential to the Euclidean localization, such that $V'(B)={\cal R}_r(B,0)$. We shall now employ this simpler setting to move away from the chiral limit.\footnote{We mention that, were we not to consider the simplifying limit $B_\star\to\infty$, certain properties would remain $B_\star$-independent, such as any property along the $T=0$ or $\mu=0$ axes. This includes the crossover temperature at $\mu=0$ or the values for $\mu_{\rm ls}$, $\mu_{\rm first}$ and $\mu_{\rm us}$ for any $H$, as well as the function $m_+(H)$ discussed below.}\\

For a non-zero bare mass, chiral symmetry is explicitly broken. As already mentioned, the second order transition line turns into a crossover and the tricritical point into a critical point. Unlike the critical endpoint, the crossover line has no unique definition. Moreover, there exist as many crossover lines as there are order parameters. In what follows, we define the crossover temperature by the inflection of $B$ as a function of the temperature. With this choice, we can determine for which value of $H$ the crossover temperature of the quark mass function becomes $T_\chi=170$ MeV in the limit of vanishing chemical potential, which is the value found by lattice simulations at the physical point for two flavors \cite{Karsch:2000kv}. Within our approach, we treat this particular value of $H$ to correspond to the ``physical point'', $H_{\rm phys}$. We can determine it conveniently as follows. Denoting by $H={\cal R}(B)$ the gap equation in both localization schemes and taking two $T$-derivatives, we have
\beq
0 & = & \frac{\partial {\cal R}}{\partial B}\frac{dB}{dT}+\frac{\partial{\cal R}}{\partial T}\,,\label{eq:eq1}\\
0 & = & \frac{\partial^2 {\cal R}}{\partial B^2}\left(\frac{dB}{dT}\right)^2\!\!\!+2\frac{\partial^2 {\cal R}}{\partial B\partial T}\frac{dB}{dT}+\frac{\partial^2{\cal R}}{\partial T^2}+\frac{\partial {\cal R}}{\partial B}\frac{d^2B}{dT^2}\,.\label{eq:eq2}\nonumber\\
\eeq
Imposing the inflection condition $d^2B/dT^2=0$ and upon plugging (\ref{eq:eq1}) into (\ref{eq:eq2}), we arrive at the following condition
\beq
0= \frac{\partial^2 {\cal R}}{\partial B^2}\left(\frac{\partial{\cal R}}{\partial T}\right)^2\!\!\!-2\frac{\partial^2 {\cal R}}{\partial B\partial T}\frac{\partial{\cal R}}{\partial B}\frac{\partial{\cal R}}{\partial T}+\frac{\partial^2 {\cal R}}{\partial T^2}\left(\frac{\partial{\cal R}}{\partial B}\right)^2,\nonumber\\
\eeq
which we can solve for $B$, given the expected crossover temperature. Knowing the crossover value of $B$, we can then determine $H_{\rm phys}$ from the gap equation. For the physical localization, we find $H_{\rm phys}=33$ MeV whereas for the Euclidean localization, we find $H_{\rm phys}=10$ MeV. Once $H_{\rm phys}$ is determined, we can then locate the critical point in the associated phase diagram, see Tab.~\ref{critphysical}. 

\begin{table}[t!]
  \centering
  	\begin{tabular}{|c ||| c || c|||} 
		\hline
		 Models for CEP &  $\mu_c^{\rm phys}$ & $T_c^{\rm phys}$ \\ [0.5ex] 
		\hline\hline	
		Physical localization & 389  & 35\\
		\hline
		Euclidean localization & 427 & 22\\
		\hline
		Hatta {\it et al.} \cite{Hatta:2002sj}  & 279 & 95\\
		\hline
		Ayala  {\it et al.} \cite{Ayala:2017ucc} & 315-349 & 18-45\\
		  \hline
		 Cui {\it et al.} \cite{Cui:2017ilj}& 245 & 38 \\
		  \hline
		 Yokota {\it et al.} \cite{Yokota:2016ovb}& 287 & 5 \\
		  \hline
		 Contrera {\it et al.} \cite{Contrera:2016rqj}& 319 & 70\\
		  \hline
		  Knaute {\it et al.} \cite{Knaute:2017opk}& 204  & 112\\
		  \hline
		 Antoniou {\it et al.} \cite{Antoniou:2017vti} & 256 & 150 \\
		  \hline
		  Scavenius {\it et al.} \cite{Scavenius:2000qd} L$\sigma$M & 207 & 99 \\
		  \hline
		 Scavenius {\it et al.} \cite{Scavenius:2000qd} NJL & 332 & 46\\
		  \hline
		 Fischer {\it et al.} \cite{Fischer:2014ata}& 168 & 115 \\
		  \hline
		 Tripolt {\it et al.} \cite{Tripolt:2014wra} & 293 & 10 \\
		  \hline
		  Costa {\it et al.} \cite{Costa:2008yh}& 332 & 80 \\
		  \hline
		  Kovacs {\it et al.} \cite{Kovacs:2007sy}& 320 & 63\\
		\hline
	\end{tabular}
	\caption{Coordinates of the critical point in the phase diagram at the ``physical point'', $H_{\rm phys}$. All values for $m=500$ and in MeV. We compare our findings with many literature model computations of the QCD CEP.}
	\label{critphysical}
\end{table}

As can be seen, the community has not yet reached a ballpark consensus on the location of the critical end point in the QCD phase diagram and a wide range of results seem permissible at this point. Our numbers do certainly fall within the group of lower temperatures and larger chemical potentials.\\

Finally, one can study how our findings for the phase diagram depend on the gluon mass of the Curci-Ferrari model. While $m=500$ MeV is the value that globally works best in both the pure Yang-Mills as well as the unquenched sector, it is nonetheless insightful to vary it as a free parameter. 
Thereby, for each value of $m$, we always insist on fixing the coupling such that we keep the $T=\mu=0$ solution $B_0$ fixed at $300$ MeV, in the chiral limit. Away from the chiral limit, we only vary $H$, without further changing $g$.

\begin{figure}[t]
\centering
	\includegraphics[width=0.45\textwidth]{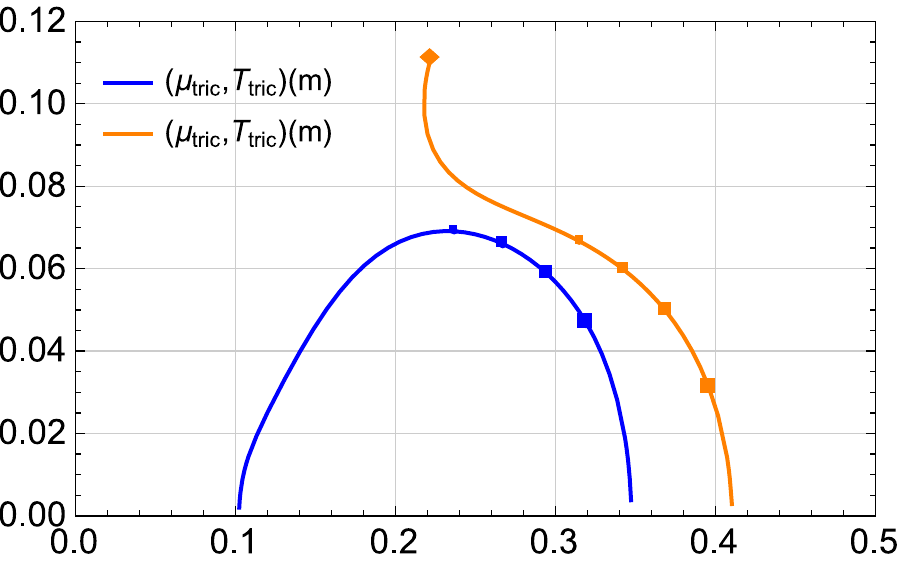}
	\caption{\label{tricofm} Position of the tricritical points upon varying the gluon mass in the two considered localizations (physical in blue, Euclidean in orange). The various points correspond to $m=x\,500$ MeV, with $x=1$, $1.25$, $1.5$ and $1.75$. The diamond point on the Euclidean localization curve corresponds to $m=0$ and its coordinates are $\mu=222$ MeV and $T=112$ MeV. At the other end of the curve, we have $\mu=410$ MeV attained for $m=937$ MeV.}
\end{figure}

In Fig. \ref{tricofm}, we display the position of the tricritical point in the chiral limit as the Curci-Ferrari mass parameter is varied. As can be seen, the obtained trajectories are qualitatively quite different depending on the considered localization scheme, although for $m=500$ MeV, the tricritical points are not so far apart, in particular in temperature values. Interestingly, while in all localization schemes considered, the gluon mass can never exceed an upper limit $m_+$, in the Euclidean localization, one might take $m\rightarrow0$ without losing the tricritical point, so $m_-=0$ in this case.

\begin{figure}[t]
\centering
	\includegraphics[width=0.4\textwidth]{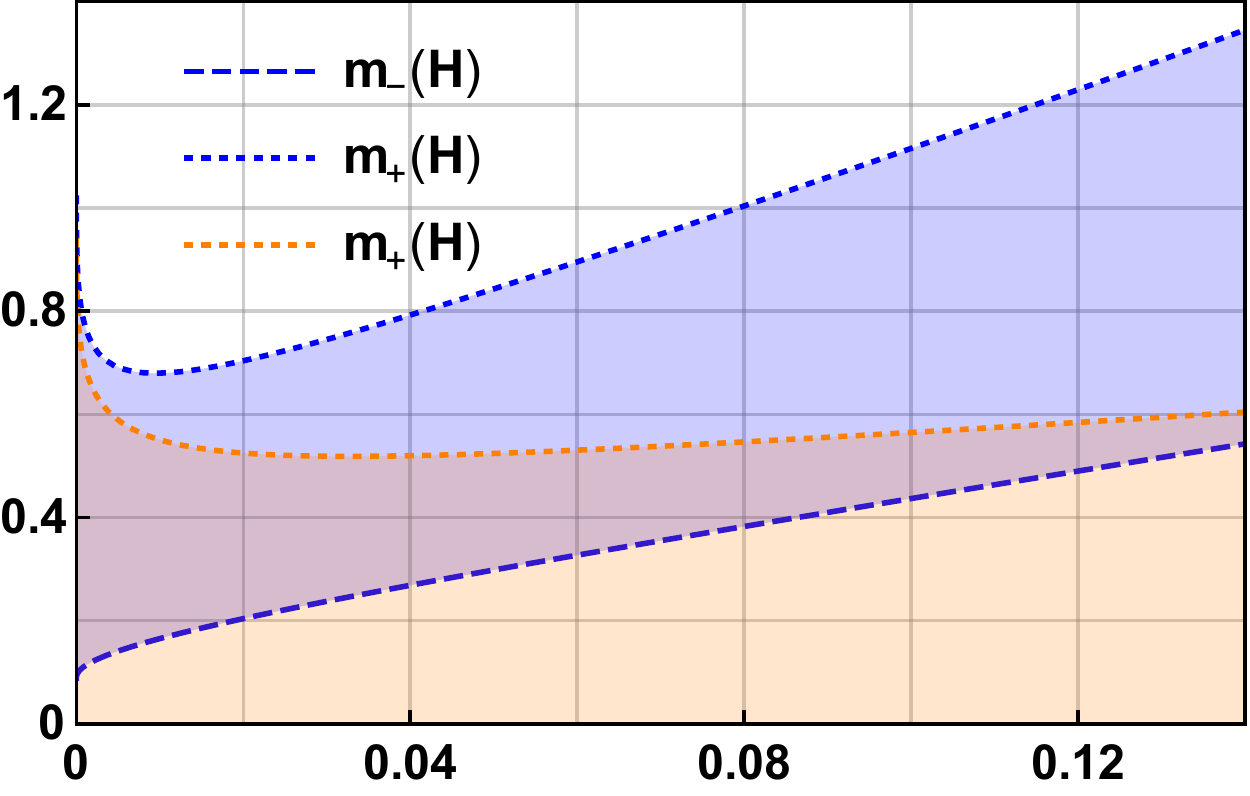}
	\caption{\label{mofH} Comparison of the functions $m_\pm(H)$ in the physical and Euclidean localizations. The respective shaded areas denote the parameter values compatible with the existence of a CEP.}
\end{figure}

As before the definition of $m_\pm$ is trivially extended to the case of non-zero $H$ , where the tricritical point is replaced by a critical end-point. We then show our results for these values in dependence of the symmetry breaking parameter $H$ in Fig.\ref{mofH} and in comparison with the findings in the physical localization. Here, we iterate our observation that the existence of a CEP puts an upper bound on the allowed for values for the gluon mass in both localization schemes considered, whereas a lower bound only exists for the physical one.

\subsection{Chiral condensate}
As mentioned previously, the mostly used order parameter for the chiral transition is not the constituent quark mass but the chiral condensate. Within the localized schemes considered here, it is natural to approximate the chiral condensate as
\beq
\sigma=-4N_cN_f B\,J_B\,,
\eeq
with
\beq
J_B & \equiv & \int_{\hat Q}^T \frac{1}{Q^2_{i\mu}+B^2}\,.
\eeq
At this level of description, we can adopt an adhoc renormalization of the condensate by removing the divergence in $J_B$, up to the scale in the logarithm of the vacuum contribution of $J_B$ which makes the renormalized condensate $\bar\sigma$ a scale dependent quantity:
\beq
\bar\sigma=-4N_cN_f B\,\bar J_B\,,
\eeq
with
\beq
\bar J_B & \equiv & -\frac{B^2}{16\pi^2}\left[\ln\frac{\bar\mu^2}{B^2}+1\right]\nonumber\\
& & +\,\frac{1}{4\pi^2}\int_0^\infty dq\,\frac{q^2}{\varepsilon^{B}_{q}}\,\Big[n^{(+)}_{\varepsilon^{B}_{q}-\mu}+n^{(+)}_{\varepsilon^{B}_{q}+\mu}\Big]\,.
\eeq
To get some intuition on the behavior of $\bar\sigma$, consider the physically localized gap equation which we rewrite identically as
\beq
H=4\pi^2 B\left[\frac{\bar J_m-\bar J_B}{m^2-B^2}-\frac{\bar J^{\rm vac}_m-\bar J^{\rm vac}_{B_0}}{m^2-B^2_0}\right].
\eeq
In the limit $T\to\infty$, the tadpole sum-integral $\bar J_m$ grows like $\sim T^2/12$, which is only counter-balanced provided $B$ vanishes as $1/T^2$, in which case we also have $\bar J_B\sim - T^2/24$. Putting all the pieces together, we find
\beq
B\sim \frac{2}{\pi^2} \frac{m^2H}{T^2}
\eeq
and thus
\beq
\bar\sigma\to \frac{N_cN_f}{3\pi^2} m^2H\,.
\eeq

\begin{figure}[t]
	\centering
	\includegraphics[width=0.42\textwidth]{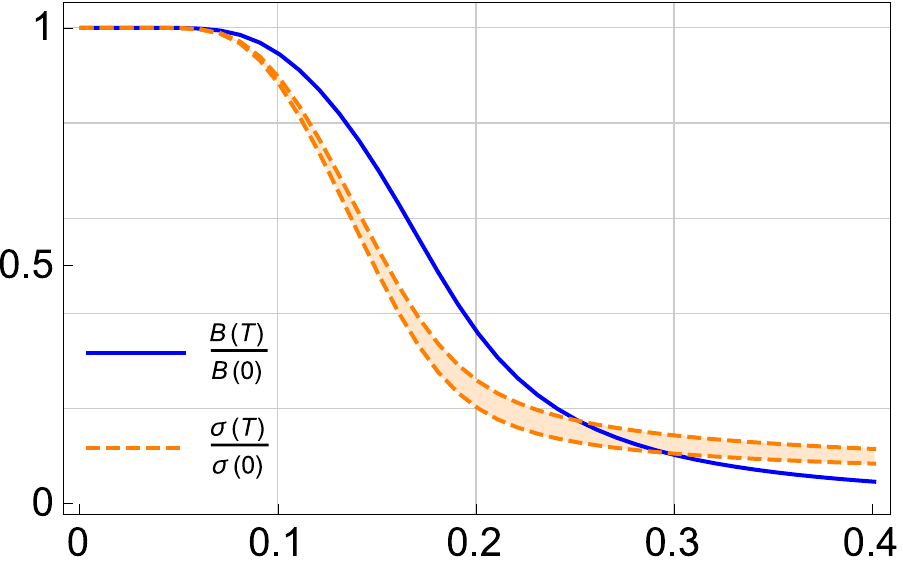}\\
	\vglue4mm
	\includegraphics[width=0.42\textwidth]{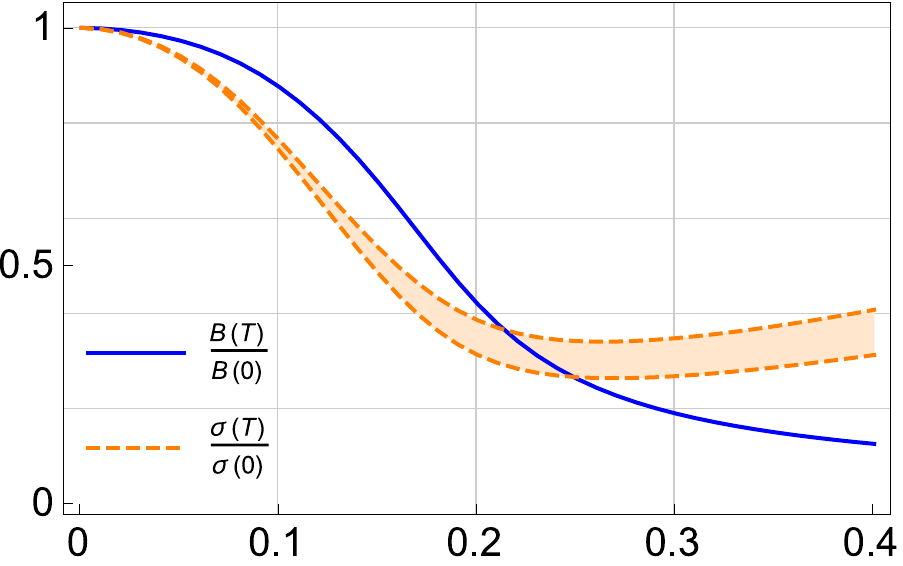}\\
	\caption{Mass and renormalized chiral condensate (top plot: physical localization; bottom plot: Euclidean localization) as functions of the temperature and at $\mu=0$. The band for the renormalized chiral condensate is obtained by varying the renormalization scale $\bar\mu$ in $\bar J_B$ by $\pm 20\%$ around $1$\,GeV. Both plots are obtained by adjusting $H$ such that the crossover temperature associated to $B$ is $170$ MeV.}
	\label{fig:cond}
\end{figure}
In the Euclidean localization the condensate is found to grow with $T^2$ at large $T$, which is clearly an artefact.\footnote{The coefficient is proportional to $H$ however in such a way that the condensate becomes zero in the chiral limit, in the high temperature region.} Despite this feature, in both cases, we can define a crossover temperature associated to the inflexion point of the chiral condensate as a function of the temperature. It is found to be lower than the corresponding crossover temperature for $B$. We can also try to fine-tune the value of $H$ to bring the crossover temperature for the condensate to the lattice value of $170$ MeV. To this purpose, consider the equation $d^2\bar\sigma/dT^2=0$. Seeing $\bar\sigma$ as a function of $B$ and $T$, this equation rewrites
\beq
0 & = & \left(\frac{\partial{\cal R}}{\partial T}\right)^2\left(\frac{\partial^2{\cal R}}{\partial B^2}\frac{\partial\sigma}{\partial B}-\frac{\partial^2\sigma}{\partial B^2}\frac{\partial{\cal R}}{\partial B}\right)\nonumber\\
& - & 2\frac{\partial{\cal R}}{\partial T}\frac{\partial{\cal R}}{\partial B}\left(\frac{\partial^2{\cal R}}{\partial B\partial T}\frac{\partial\sigma}{\partial B}-\frac{\partial^2\sigma}{\partial B\partial T}\frac{\partial{\cal R}}{\partial B}\right)\nonumber\\
& + & \left(\frac{\partial{\cal R}}{\partial B}\right)^2\left(\frac{\partial^2{\cal R}}{\partial T^2}\frac{\partial\sigma}{\partial B}-\frac{\partial^2\sigma}{\partial T^2}\frac{\partial{\cal R}}{\partial B}\right),
\eeq
which we use again to determine the crossover value for $B$, assuming a crossover temperature $T_\chi=170$\,MeV. The corresponding value of $H$ is then obtained from the gap equation. In the physical localization, we find $H_{\rm phys}=117$\, MeV, and a critical end-point located at $(504\,{\rm MeV},11\,{\rm MeV})$, but the value of $B$ gets suspiciously close to the bound $B=m$. In the Euclidean localization, it seems not to be possible to reach these transition temperatures, at least at this level of approximation.

\section{Conclusion}

We have computed the phase diagram of QCD with light quarks in the context of a first-principle inspired approach to infrared QCD, the Curci-Ferrari model, using the double expansion in the pure gauge coupling and in $1/N_c$ proposed in Ref.~\cite{Pelaez:2017bhh}. To allow for a semi-analytic grasp of the corresponding equations and since our first aim is a qualitative survey of what to expect when the equations are solved in full glory, we have used some simplifying approximations, in particular the localization scheme discussed in \cite{Reinosa:2011cs,Marko:2015gpa}, which we have extended to the present context. For the parameters used here, the leading-order results agree well with those of effective quark-meson models when the chiral anomaly is neglected \cite{Resch:2017vjs,Jakovac:2003ar}. Although subleading in $1/N_c$, the latter and meson fluctuations are important to correctly determine the phase structure in the Columbia plot \cite{Pisarski:1983ms,Lenaghan:2000kr,Resch:2017vjs}.  In principle, they can be systematically included at next-to-leading order in the present expansion scheme. 

We also mention that similar rainbow equations have already been considered in great detail to study the phase diagram in the context of nonperturbative functional approaches \cite{Fischer:2012vc,Hatta:2002sj}. Not too surprisingly, our results agree qualitatively with those (the main differences may be attributed to the different treatments of the gluon sector). However, the key point of the present work is to systematically justify the employed approximation on the basis of identified small parameters in QCD. 

Although we used here simplified versions of the complete rainbow equation \eqn{eq:rainbow}, we expect, inspired by the vacuum case, that our main results are robust. In particular, we have tested that our results do not depend much on the type of localization we use. One notable exception is the fate of the (tri)critical point as the gluon mass is taken to zero. In some scenarios, the existence of a critical end-point seems to require a nonzero gluon mass. It would be interesting to investigate to what extent this is an artefact of the localization procedure by solving the original equation (\ref{RB1}). Another, obvious extension of the present work will be to include the other scalar functions $A_0$, $A_v$, and $C$ in \Eqn{eq:rainbow}, see Refs.~\cite{Roberts:2000aa,Fischer:2018sdj}.

Yet another interesting direction of investigation is to include the order parameter of the deconfinement transition, the Polyakov loop, in the spirit of Refs.~\cite{Reinosa:2014ooa,Reinosa:2015oua,Maelger:2017amh}, which would allow one to study the interplay between the chiral and deconfinement phase transition across the Columbia plot \cite{Fukushima:2003fw,Schaefer:2007pw,Folkestad:2018psc}. 

\acknowledgements{We thank Zs. Sz\'ep for interesting discussions and helpful remarks concerning the manuscript.}

\appendix

\section{Symmetries of the quark propagator}\label{app:symmetries}
The full propagator $S$ is a function of the external momentum variables $\omega$ and $\vec p$ as well as $\mu$ and $T$. Dropping the explicit $T$-dependence for notational simplicity, and assuming isotropy, its tensor structure can decomposed as
\beq
S(\omega,\vec p; \mu)&=& S_1\mathds{1}+S_2\gamma_5+S_3\gamma_0+S_4\gamma_0\gamma_5
\nonumber\\
&& +\,S_5\, \hat p\cdot \vec \gamma +S_6\, \hat p\cdot \vec \gamma \gamma_5+S_7\,\gamma_0\hat p\cdot \vec \gamma \,,
\eeq
where $S_i=S_i(\omega, p; \mu)$, $p\equiv |\vec{p}|$ and $\hat p\equiv \vec{p}/p$. Under a parity transformation 
\beq
S(\omega,\vec p; \mu) &\rightarrow& \gamma_0 S(\omega,-\vec p; \mu) \gamma_0 
\nonumber\\
&=&
S_1\mathds{1}-S_2\gamma_5+S_3\gamma_0-S_4\gamma_0\gamma_5
\nonumber\\
&& +\,S_5\, \hat p\cdot \vec \gamma -S_6\, \hat p\cdot \vec \gamma \gamma_5+S_7\,\gamma_0\hat p\cdot \vec \gamma \,,
\eeq
and thus parity invariance implies $S_2=S_4=S_6=0$. In fact, in what follows, it is convenient to use the parametrisation
\beq
S(\omega,\vec p; \mu)=\tilde B\mathds{1}+(i\omega-\mu)\gamma_0 \tilde A_0+i\vec p\cdot \vec \gamma \tilde A_v+i\gamma_0\vec p\cdot \vec \gamma \,\tilde C\,.\nonumber\\
\eeq
Under charge conjugation 
\beq
& & S(\omega,\vec p; \mu)\rightarrow\gamma_2\gamma_0 S(-\omega,-\vec p; -\mu)^{\rm t} \gamma_0 \gamma_2
\nonumber\\
& & \hspace{0.4cm}=\,\tilde B(-\omega, p; -\mu)\mathds{1}  +(i\omega-\mu)\gamma_0\tilde A_0(-\omega, p; -\mu)
\nonumber\\
& & \hspace{0.4cm}+\,i\vec p\cdot \vec \gamma  \tilde A_v(-\omega, p; -\mu)+i\gamma_0\vec p\cdot \vec \gamma \,\tilde C(-\omega, p; -\mu)\,,\nonumber\\
\eeq
where we have used $\gamma_2\gamma_0\gamma_\mu^{\rm t}\gamma_0\gamma_2=-\gamma_\mu$, valid in the particular (Weyl) representation of the $\gamma_\mu$ matrices considered here. Charge conjugation invariance then implies
\beq\label{eq:C}
\tilde X(-\omega, p; -\mu)=\tilde X(\omega, p; \mu)\,,
\eeq
for any of the components $\tilde X=\tilde A_0,\tilde A_v,\tilde B,\tilde C$. Similarly, under complex conjugation
\beq
& & S(\omega,\vec p; \mu)\rightarrow\gamma_3\gamma_1 S(-\omega,-\vec p; \mu^{*})^{*} \gamma_1 \gamma_3
\nonumber\\
& &\hspace{0.4cm}=\,\tilde B(-\omega, p; \mu^{*})^{*}\mathds{1}  +(i\omega-\mu)\gamma_0\tilde A_0(-\omega, p; \mu^{*})^{*}
\nonumber\\
& & \hspace{0.4cm}+\,i\vec p\cdot \vec \gamma \tilde A_v(-\omega, p; \mu^{*})^{*}+i\gamma_0\, \vec p\cdot \vec \gamma \,\tilde C(-\omega, p; \mu^{*})^{*}\,, \nonumber\\
\eeq
where we have used $\gamma_3\gamma_1\gamma_\mu^{*}\gamma_1\gamma_3=\gamma_\mu$, valid, again, in the Weyl representation. It follows that
\beq\label{eq:K}
\tilde X(-\omega, p; \mu^*)^*=\tilde X(\omega, p; \mu)\,,
\eeq
for any of the components $\tilde X=\tilde A_0,\tilde A_v,\tilde B,\tilde C$. Combining (\ref{eq:C}) and (\ref{eq:K}), we also obtain
\beq\label{eq:CK}
\tilde X(\omega, p; -\mu^*)^*=\tilde X(\omega, p; \mu)\,.
\eeq
In particular, all components are real in the case of an imaginary chemical potential. In the case of a real chemical potential, these components become complex, the real and imaginary parts, corresponding to the frequency even and odd parts, $(\tilde X(\omega,p;\mu)+\tilde X(-\omega,p;\mu))/2$ and $(\tilde X(\omega,p;\mu)-\tilde X(-\omega,p;\mu))/2i$, respectively.

\section{Retarded Green's function at finite $\mu$}\label{app:mu_prescription}

We briefly recall the origin of \Eqn{eq:ret_prescription}, considering, for simplicity, the case of a charged scalar field. The  physical retarded propagator is defined as
\beq
\smash{G^{\rm phys}_{\rm ret}(t)\equiv-i\frac{\Theta(t)}{Z}{\rm tr}\,e^{-\beta (H+\mu Q)}\,[\varphi_H(t),\varphi^\dagger(0)}],
\eeq
where $\smash{Z\equiv {\rm tr}\,e^{-\beta (H+\mu Q)}}$ denotes the grand-canonical partition function, and $\smash{\varphi_H(t)=e^{iHt}\varphi(0)e^{-iHt}}$ is the Heisenberg field, evolving according to $H$ and not $H+\mu Q$. Let us note that we use an unconventional sign for the chemical potential (to be consistent with our choice in \cite{Reinosa:2015oua,Maelger:2017amh}) while keeping the usual convention for the charge, such that $[Q,\varphi]=-\varphi$ and $[Q,\varphi^\dagger]=\varphi^\dagger$. 

Now, we would like to relate the physical retarded propagator to the Matsubara propagator defined as ($\smash{0<\tau<\beta}$)
\beq
G_{\rm Mat}(\tau)\equiv\frac{1}{Z}{\rm tr}\,e^{-\beta (H+\mu Q)}\,\varphi_{H+\mu Q}(-i\tau)\varphi^\dagger(0)\,,
\eeq
where now the Heisenberg field evolves in imaginary time, according to $H+\mu Q$, not $H$. The reason for defining the Matsubara propagator in this way is that it possesses a simple functional integral representation. 

The relation between the two propagators $G^{\rm phys}_{\rm ret}(t)$ and $G_{\rm Mat}(\tau)$ is now most easily derived by introducing an auxiliary retarded propagator 
\beq
G_{\rm ret}(t)\!\equiv\!-i\frac{\Theta(t)}{Z}{\rm tr}\,e^{-\beta (H+\mu Q)}[\varphi_{H+\mu Q}(t),\varphi^\dagger(0)].
\eeq  
Inserting a complete basis of states under the trace, this relation in Fourier space is found to be 
\beq\label{eq:123}
G_{\rm ret}(\omega)=G_{\rm Mat}(\omega_n\to -i\omega+0^+)\,.
\eeq 
Moreover, from the commutators of $Q$ with $\varphi$ or $\varphi^\dagger$, one obtains 
\beq
G^{\rm phys}_{\rm ret}(t)=e^{i\mu t}G_{\rm ret}(t)\,,
\eeq 
and thus 
\beq\label{eq:456}
G^{\rm phys}_{\rm ret}(\omega)=G_{\rm ret}(\omega+\mu)\,.
\eeq
Combining (\ref{eq:123}) and (\ref{eq:456}), we arrive at the desired result
\beq
G^{\rm phys}_{\rm ret}(\omega)=G_{\rm Mat}(\omega_n\to -i(\omega+\mu)+0^+)\,.
\eeq


\begin{thebibliography}{}

\bibitem{Fukushima:2010bq}
  K.~Fukushima and T.~Hatsuda,
  Rept.\ Prog.\ Phys.\  {\bf 74} (2011) 014001.

\bibitem{Weissenborn:2011qu}
  S.~Weissenborn, I.~Sagert, G.~Pagliara, M.~Hempel and J.~Schaffner-Bielich,
  Astrophys.\ J.\  {\bf 740} (2011) L14.
  
\bibitem{Borsanyi:2016ksw}
  S.~Bors\'anyi {\it et al.},
  Nature (London) {\bf 539} (2016)  69.

\bibitem{Aoki:2006br}
  Y.~Aoki, Z.~Fodor, S.~D.~Katz and K.~K.~Szabo,
  Phys.\ Lett.\ B {\bf 643} (2006) 46.

\bibitem{Bazavov:2011nk}
  A.~Bazavov {\it et al.},
  Phys.\ Rev.\ D {\bf 85} (2012) 054503.

\bibitem{DElia:2018fjp}
  M.~D'Elia,
  Nucl.\ Phys.\ A {\bf 982} (2019) 99.

\bibitem{Pisarski:1983ms}
  R.~D.~Pisarski and F.~Wilczek,
  Phys.\ Rev.\ D {\bf 29} (1984) 338.

\bibitem{deForcrand:2010ys}
  P.~de Forcrand,
  PoS LAT {\bf 2009} (2009) 010.


  
\bibitem{Stephanov:2004wx}
  M.~A.~Stephanov,
  Prog.\ Theor.\ Phys.\ Suppl.\  {\bf 153} (2004) 139
   [Int.\ J.\ Mod.\ Phys.\ A {\bf 20} (2005) 4387].
  

\bibitem{Mohanty:2005mv}
  B.~Mohanty and J.~Serreau,
  Phys.\ Rept.\  {\bf 414} (2005) 263.

\bibitem{Stephanov:2008qz}
  M.~A.~Stephanov,
  Phys.\ Rev.\ Lett.\  {\bf 102} (2009) 032301.
  
\bibitem{Hatta:2002sj}
  Y.~Hatta and T.~Ikeda,
  Phys.\ Rev.\ D {\bf 67} (2003) 014028.

\bibitem{Roberts:2000aa}
  C.~D.~Roberts and S.~M.~Schmidt,
  Prog.\ Part.\ Nucl.\ Phys.\  {\bf 45} (2000) S1.

\bibitem{Fischer:2018sdj} 
  C.~S.~Fischer,
  Prog.\ Part.\ Nucl.\ Phys.\  {\bf 105}, 1 (2019).

\bibitem{Fukushima:2013rx}
  K.~Fukushima and C.~Sasaki,
  Prog.\ Part.\ Nucl.\ Phys.\  {\bf 72} (2013) 99.

\bibitem{Ding:2015ona}
  H.~T.~Ding, F.~Karsch and S.~Mukherjee,
  Int.\ J.\ Mod.\ Phys.\ E {\bf 24} (2015) 1530007.
   
\bibitem{Mohanty:2011nm}
  B.~Mohanty (STAR Collaboration),
  J.\ Phys.\ G {\bf 38} (2011) 124023.

\bibitem{Senger:2016wfb}
  P.~Senger,
  Eur.\ Phys.\ J.\ A {\bf 52} (2016) 217.
  
\bibitem{Ablyazimov:2017guv}
  T.~Ablyazimov {\it et al.} [CBM Collaboration],
  Eur.\ Phys.\ J.\ A {\bf 53} (2017)  60.

\bibitem{Sakaguchi:2017ggo}
  T.~Sakaguchi [J-PARC-HI Collaboration],
  Nucl.\ Phys.\ A {\bf 967} (2017) 896.

\bibitem{Fodor:2018wul}
  Z.~Fodor {\it et al.},
  arXiv:1807.09862 [hep-lat].

\bibitem{Fischer:2012vc}
  C.~S.~Fischer and J.~Luecker,
  Phys.\ Lett.\ B {\bf 718} (2013) 1036.

\bibitem{Fischer:2013eca}
  C.~S.~Fischer, L.~Fister, J.~Luecker and J.~M.~Pawlowski,
  Phys.\ Lett.\ B {\bf 732} (2014) 273.

\bibitem{Eichmann:2015kfa}
  G.~Eichmann, C.~S.~Fischer and C.~A.~Welzbacher,
  Phys.\ Rev.\ D {\bf 93} (2016)  034013.

\bibitem{Jakovac:2003ar}
  A.~Jakov\'ac, A.~Patk\'os, Z.~Sz\'ep and P.~Sz\'epfalusy,
  Phys.\ Lett.\ B {\bf 582} (2004) 179;
  Acta Phys.\ Hung.\ A {\bf 22} (2005) 355.

\bibitem{Schaefer:2007pw}
  B.~J.~Schaefer, J.~M.~Pawlowski and J.~Wambach,
  Phys.\ Rev.\ D {\bf 76} (2007) 074023.
  
\bibitem{Fukushima:2008wg}
  K.~Fukushima,
  Phys.\ Rev.\ D {\bf 77} (2008) 114028
   Erratum: [Phys.\ Rev.\ D {\bf 78} (2008) 039902].

\bibitem{Herbst:2010rf}
  T.~K.~Herbst, J.~M.~Pawlowski and B.~J.~Schaefer,
  Phys.\ Lett.\ B {\bf 696} (2011) 58;
  Phys.\ Rev.\ D {\bf 88} (2013) 014007

\bibitem{Resch:2017vjs} 
  S.~Resch, F.~Rennecke and B.~J.~Schaefer,
  Phys.\ Rev.\ D {\bf 99}, no. 7, 076005 (2019).
  
\bibitem{Curci:1976bt}
  G.~Curci and R.~Ferrari,
  Nuovo Cim.\ A {\bf 32} (1976) 151.

\bibitem{Tissier:2010ts}
  M.~Tissier and N.~Wschebor,
  Phys.\ Rev.\ D {\bf 82} (2010) 101701;
  Phys.\ Rev.\ D {\bf 84} (2011) 045018.

\bibitem{Bogolubsky:2009dc}
  I.~L.~Bogolubsky, E.~M.~Ilgenfritz, M.~Muller-Preussker and A.~Sternbeck,
  Phys.\ Lett.\ B {\bf 676} (2009) 69.

\bibitem{Serreau:2012cg}
  J.~Serreau and M.~Tissier,
  Phys.\ Lett.\ B {\bf 712} (2012) 97.

\bibitem{Reinosa:2017qtf}
  U.~Reinosa, J.~Serreau, M.~Tissier and N.~Wschebor,
  Phys.\ Rev.\ D {\bf 96} (2017) 014005.

\bibitem{Reinosa:2014ooa}
  U.~Reinosa, J.~Serreau, M.~Tissier and N.~Wschebor,
  Phys.\ Lett.\ B {\bf 742} (2015) 61;
  Phys.\ Rev.\ D {\bf 91} (2015) 045035;
  Phys.\ Rev.\ D {\bf 93} (2016) 105002.

\bibitem{Reinosa:2015oua}
  U.~Reinosa, J.~Serreau and M.~Tissier,
  Phys.\ Rev.\ D {\bf 92} (2015) 025021.

\bibitem{Maelger:2017amh}
  J.~Maelger, U.~Reinosa and J.~Serreau,
  Phys.\ Rev.\ D {\bf 97} (2018) 074027.
  Phys.\ Rev.\ D {\bf 98} (2018) 094020.

\bibitem{Skullerud:2003qu}
  J.~I.~Skullerud, P.~O.~Bowman, A.~Kizilersu, D.~B.~Leinweber and A.~G.~Williams,
  JHEP {\bf 0304} (2003) 047.

\bibitem{Pelaez:2017bhh}
  M.~Pel\'aez, U.~Reinosa, J.~Serreau, M.~Tissier and N.~Wschebor,
  Phys.\ Rev.\ D {\bf 96} (2017)  114011.
  
\bibitem{Reinosa:2011cs} 
  U.~Reinosa and Zs.~Sz\'ep,
  Phys.\ Rev.\ D {\bf 85}, 045034 (2012).
  
\bibitem{Marko:2015gpa} 
  G.~Mark\'o, U.~Reinosa and Zs.~Sz\'ep,
  Phys.\ Rev.\ D {\bf 92} (2015)125035.
  
\bibitem{Laine:2016hma} 
  M.~Laine and A.~Vuorinen,
  Lect.\ Notes Phys.\  {\bf 925}, pp.1 (2016).
  
\bibitem{Cohen:2003kd}
  T.~D.~Cohen,
  Phys.\ Rev.\ Lett.\  {\bf 91} (2003) 222001.

\bibitem{Marko:2014hea} 
  G.~Mark\'o, U.~Reinosa and Zs.~Sz\'ep,
  Phys.\ Rev.\ D {\bf 90} (2014) 125021.

\bibitem{RILO_bg}
J.~Maelger, U.~Reinosa and J.~Serreau, in preparation. 


\bibitem{Schaefer:2004en} 
  B.~J.~Schaefer and J.~Wambach,
  Nucl.\ Phys.\ A {\bf 757}, 479 (2005).

\bibitem{Qin:2010nq} 
  S.~x.~Qin, L.~Chang, H.~Chen, Y.~x.~Liu and C.~D.~Roberts,
  Phys.\ Rev.\ Lett.\  {\bf 106}, 172301 (2011).
  
\bibitem{Costa:2008yh} 
  P.~Costa, M.~C.~Ruivo and C.~A.~de Sousa,
  Phys.\ Rev.\ D {\bf 77}, 096001 (2008).
  
\bibitem{Ayala:2017ucc} 
  A.~Ayala, S.~Hernandez-Ortiz and L.~A.~Hernandez,
  Rev.\ Mex.\ Fis.\  {\bf 64}, no. 3, 302 (2018).
  
\bibitem{Cui:2017ilj} 
  Z.~F.~Cui, J.~L.~Zhang and H.~S.~Zong,
  Sci.\ Rep.\  {\bf 7}, 45937 (2017).
  
\bibitem{Yokota:2016ovb} 
  T.~Yokota, T.~Kunihiro and K.~Morita,
  arXiv:1611.06669 [hep-ph].
  
\bibitem{Contrera:2016rqj} 
  G.~A.~Contrera, A.~G.~Grunfeld and D.~Blaschke,
  Eur.\ Phys.\ J.\ A {\bf 52}, no. 8, 231 (2016).
  
\bibitem{Knaute:2017opk} 
  J.~Knaute, R.~Yaresko and B.~K\"ampfer,
  Phys.\ Lett.\ B {\bf 778}, 419 (2018).
  
\bibitem{Antoniou:2017vti} 
  N.~G.~Antoniou, F.~K.~Diakonos, X.~N.~Maintas and C.~E.~Tsagkarakis,
  Phys.\ Rev.\ D {\bf 97}, no. 3, 034015 (2018).
  
\bibitem{Fischer:2014ata} 
  C.~S.~Fischer, J.~Luecker and C.~A.~Welzbacher,
  Phys.\ Rev.\ D {\bf 90}, no. 3, 034022 (2014).
  
\bibitem{Scavenius:2000qd} 
  O.~Scavenius, A.~Mocsy, I.~N.~Mishustin and D.~H.~Rischke,
  Phys.\ Rev.\ C {\bf 64}, 045202 (2001).
  
\bibitem{Tripolt:2014wra} 
  R.~A.~Tripolt, L.~von Smekal and J.~Wambach,
  Phys.\ Rev.\ D {\bf 90}, no. 7, 074031 (2014).
  
\bibitem{Kovacs:2007sy} 
  P.~Kovacs and Z.~Szep,
  Phys.\ Rev.\ D {\bf 77}, 065016 (2008).
  
\bibitem{Karsch:2000kv} 
  F.~Karsch, E.~Laermann and A.~Peikert,
  Nucl.\ Phys.\ B {\bf 605}, 579 (2001).
  
\bibitem{Lenaghan:2000kr}
  J.~T.~Lenaghan,
  Phys.\ Rev.\ D {\bf 63} (2001) 037901.

\bibitem{Fukushima:2003fw}
  K.~Fukushima,
  Phys.\ Lett.\ B {\bf 591} (2004) 277.

\bibitem{Folkestad:2018psc} 
  A.~Folkestad and J.~O.~Andersen,
  Phys.\ Rev.\ D {\bf 99}, no. 5, 054006 (2019).

\end{thebibliography}
\end{document}